\documentclass[aps,prd,superscriptaddress,nofootinbib,11pt]{revtex4}
\usepackage[english]{babel}
\usepackage[utf8]{inputenc}
\usepackage{graphicx}   % need for figures
\usepackage{slashed}
\usepackage{epstopdf}
\usepackage{verbatim}   % useful for program listings
\usepackage{color}      % use if color is used in text
\usepackage{subfigure}  % use for side-by-side figures
\usepackage{multirow}
\usepackage{hyperref}   % use for hypertext links, including those to external documents and URLs
\usepackage{float}
\usepackage{epsfig,rotating}
\usepackage{amsmath,amssymb}
\usepackage{dsfont}
\usepackage{slashed}
\restylefloat{table}
\numberwithin{equation}{section}

\newcommand{\vx}{\vec{x}}
\newcommand{\vp}{\vec{p}}
\newcommand{\vq}{\vec{q}}
\newcommand{\vk}{\vec{k}}

\newcommand{\be}{\begin{equation}}
\newcommand{\ee}{\end{equation}}
\newcommand{\bea}{\begin{eqnarray}}
\newcommand{\eea}{\end{eqnarray}}

\newcommand{\ket}[1]{|#1\rangle}

\newcommand{\Ok}{\ensuremath{\Omega_k}}

\begin{document}
%\tableofcontents
\title{Thermalization by off-shell processes: the virtues of small virtuality. }

\author{Daniel Boyanovsky}
\email{boyan@pitt.edu} \affiliation{Department of Physics and
Astronomy, University of Pittsburgh, Pittsburgh, PA 15260}

 \date{\today}

\begin{abstract}
 We study the thermalization   of a scalar field $\Phi$ coupled to two other scalar fields $\chi_{1,2}$ that constitute a bath in thermal equilibrium. For a range of masses the $\Phi$ propagator features threshold and infrared divergences, a vanishing residue at the (quasi) particle pole and vanishing \emph{on-shell} decay rates thereby preventing the equilibration of $\Phi$ with the bath via on-shell processes. Inspired by the theory of quantum open systems we obtain a quantum master equation for the reduced density matrix of $\Phi$ that includes the time dependence of   bath correlations, yielding time dependent rates in the dynamics of relaxation and allowing virtual processes with an energy $q_0$ whose difference  from the on-shell energy is of  $\mathcal{O}(1/t)$ at long time $t$. These \emph{off-shell} processes lead to thermalization despite vanishing S-matrix rates. In the case of threshold divergences we find that a thermal fixed point is approached as $e^{-\sqrt{t/t^*}}$ with the relaxation time $t^*$ becoming shorter at high temperature as a consequence of stimulated emission and absorption. In the infrared case, the thermal fixed point is approached as $e^{-\gamma(t)}$, where $\gamma(t)$ features a crossover between  $\ln(t)$ and  $  t$ behavior for $t \gg 1/T$. The vanishing of the residue and   the crossover in relaxational dynamics in this case is strikingly reminiscent of the orthogonality catastrophe in heavy impurity systems. The results yield more general lessons on thermalization via virtual processes.

\end{abstract}

\keywords{}

\maketitle

\section{Introduction}
The dynamics of relaxation, approach to equilibrium and thermalization,  is a subject  of timely cross disciplinary interest with implications   in condensed matter physics\cite{polko,gogo,weiss,rammer},   cosmology\cite{calzetta,bernstein,kolb,dodelson} and high energy and nuclear physics\cite{raju,blaizot,biondi}. In quantum kinetic  approaches to relaxation or thermal equilibration in non-equilibrium situations, usually the rate equations input collision kernels with transition probabilities per unit time derived from the S-matrix approach, or Fermi's golden rule, which  typically take the infinite time limit. This limit entails strict energy conservation, therefore the transition probabilities per unit time are obtained from on-shell processes that conserve energy (and momentum in translational  invariant systems).

More recently, relaxational dynamics associated with \emph{off shell effects} were studied within the context of quantum field theories that feature infrared\cite{irboyrai} and threshold divergences\cite{irthres} at zero temperature. Infrared divergences typically originate in the emission and absorption of soft quanta and are of particular importance in gauge theories\cite{bn,lee,chung,kino,kibble,yennie,weinberg,kulish,greco,schwartz1,finites,schwartz2}    and  play a fundamental role in quantum aspects of   gravity as a consequence of emission and absorption of gravitons\cite{strominger1,strominger2}. Motivated by Higgs physics, studies in refs.\cite{kni,will,thres1,thres} recognized a singularity in the propagator of a bosonic particle as its mass approaches the multiparticle threshold from below.  An important consequence and common aspect of infrared and threshold divergences is that the residue of the single particle pole vanishes as the mass of the particle approaches the multiparticle threshold, and as a consequence, the particle is not an asymptotic state of the S-matrix\cite{irthres}. Within the context of condensed matter physics, this phenomenon signals the breakdown of the quasiparticle picture\cite{orto,orto2,levitov}. In ref.\cite{irboyrai} a dynamical resummation method that extends the dynamical renormalization group\cite{gold,boyvega} was implemented to study the time evolution of an initial single particle state in the case of a bosonic quantum field theory that features an infrared divergence similar to that of the electron propagator in Quantum Electrodynamics\cite{bn,lee,chung,kino,kibble,yennie,weinberg,kulish,greco}. This study found that the survival probability of the single particle state decays in time as a consequence of off-shell, in other words virtual processes not as an exponential but as a power law with anomalous dimension, namely $t^{-\Delta}, \Delta >0$, despite the fact that the S-matrix decay rate vanishes. Implications of this off-shell process for the production of ultra light dark matter or dark radiation in a radiation dominated cosmology were studied in ref.\cite{dark}.

In ref. \cite{irthres} a similar study   showed that in the case of threshold divergences the survival probability of an initial single particle state decays as $e^{-\sqrt{t/t^*}}$ when the mass of the particle coincides with a two particle threshold for intermediate states with two massive particles, however, the decay rate obtained from the S-matrix approach vanishes in this case also. The vanishing of the residue at the single particle pole in the propagator in both the threshold and infrared divergent cases is merely a reflection of this decay process when the mass of the ``decaying'' particle places the pole at the tip of the multiparticle continuum (the beginning of the branch cut) instead of being fully embedded in the continuum as in the case of a resonance. This feature also signals a breakdown of the Breit-Wigner approximation of the propagator and spectral density of this particle\cite{irthres}.

\vspace{1mm}

\textbf{Motivation and objectives:}

The analysis in refs.\cite{irboyrai,irthres} shows that the decay of the single particle survival probability is a consequence of off shell processes which at long time $t$ feature ``virtuality'' $\propto 1/t$. While the S-matrix decay rate vanishes by strict energy (and momentum) conservation, a long but finite time interval introduces an energy uncertainty that allows processes with small virtuality that lead to the decay but with different decay law as compared to the usual exponential $e^{-\Gamma t}$. These results motivate us to address the following questions:   i) how to implement a quantum kinetic formulation that would allow time dependent rates without implementing Fermi's golden rule or without the input of S-matrix transition probabilities per unit time? ii) do these processes lead to thermalization if the particle is coupled to a thermal bath of particles with mass spectra that yield threshold and infrared divergences?.

Addressing these two questions defines our objectives in this study, namely:   \textbf{i:)} to derive and implement a quantum kinetic formulation that naturally includes time dependent rates and generalizes the dynamical resummation framework of refs.\cite{irboyrai,irthres} to the realm of finite temperature, such formulation could prove very useful in cosmology,  \textbf{ii:)} to implement this formulation to study the approach to equilibration and thermalization via off-shell processes in the cases in which the mass spectrum of particles in the thermal bath to which the particle couples yield threshold and infrared divergences.

\vspace{1mm}

\textbf{Summary of results:} We study a model of a scalar field $\Phi$ described by   an initial density matrix out of equilibrium, coupled to two other scalar fields $\chi_{1,2}$, taken to describe a bath in thermal equilibrium. By adjusting the masses, we   investigate the cases corresponding to threshold and infrared divergences, thereby allowing us to draw more general conclusions from this model.

 Inspired by the theory of open quantum systems\cite{breuer,zoeller}, we derive a quantum master equation for the reduced density matrix for the $\Phi$ field which, however, does not take the infinite time limit in the Hamiltonian nor in   the dissipation terms, thereby allowing time dependent rates in the dynamics of relaxation and off-shell processes with small virtuality $\propto 1/t$. We argue that this master equation is the generalization of the dynamical resummation method of refs.\cite{irboyrai,irthres} adapted to  describe the coupling to a thermal bath and provides a real time resummation of self-energy contributions including off-shell processes at finite temperature. In the case of threshold divergences, we find that at long time the reduced density matrix for the $\Phi$ field approaches a thermal fixed point as $e^{-\sqrt{t/t^*}}$ where the relaxation time $t^*$ shortens at high temperature as a consequence of stimulated absorption and emission. In the case of infrared divergences we find that, again, the reduced density matrix of the $\Phi$ field approaches a thermal fixed point as $e^{-\gamma(t)}$ where $\gamma(t)$ features a crossover between a $\ln(t)$ and a $\propto t$ behavior at a time scale $\propto 1/T$. The behavior $\propto t$ is a subtle consequence of infrared enhancement at finite temperature and small virtuality. The crossover time scale  and the time scale  towards thermalization increases when the $\Phi$ particle becomes relativistic.

 Remarkably, the crossover between the $\ln(t)$ and $\propto t$  behavior is strikingly similar to the orthogonality catastrophe in heavy impurity systems\cite{orto,orto2}. Off-shell effects associated with infrared singularities have   recently been studied within the context of photoexcitation of soft electron-hole pairs in graphene\cite{levitov}, hence the results of this study may prove useful to study thermalization in this system.

 Therefore, above and beyond the particular model studied here, the results obtained in this study may prove useful to study thermalization in a wide range of settings where virtual processes may play a fundamental role in relaxation an thermalization.

In section (\ref{irthres}) we introduce the model and briefly summarize the emergence of threshold and infrared divergences at zero temperature for consistency of presentation. In section (\ref{sec:master}) we derive the quantum master equation in Lindblad form\cite{breuer,zoeller,lin,gori,pearle}, discussing in detail the main assumptions but without taking the infinite time limit in the bath correlation functions. Keeping a finite time interval allows the rates in the quantum master equation to depend explicitly on time, thereby including off-shell processes of small virtuality at long time.  Furthermore, analyzing the bath correlations we establish a correspondence with the dynamical resummation framework of refs.\cite{irboyrai,irthres} and argue that the quantum master equation provides a real time resummation of self-energy corrections including bath correlations. In section (\ref{sec:cases}) we study in detail the cases of threshold and infrared divergences and obtain the main results in this article. Section (\ref{sec:discussion}) discusses potential implications of the results along with possible caveats. Section (\ref{sec:conclusion}) summarizes our conclusions and suggests further avenues of study.  Various appendices contain technical details.

\section{Threshold and infrared singularities at zero temperature:}\label{irthres}

\begin{figure}[ht!]
\begin{center}
\includegraphics[height=2.5in,width=2.5in,keepaspectratio=true]{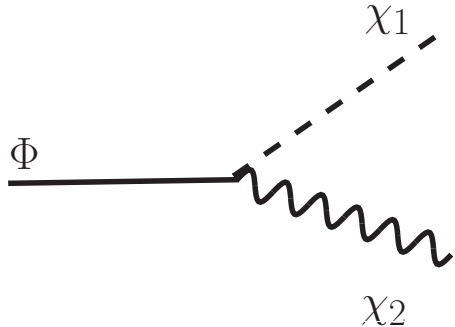}
\caption{The   interaction vertex.}
\label{fig:vertex}
\end{center}
\end{figure}

In this section we briefly summarize the main aspects of threshold and infrared divergences at zero temperature discussed in refs.\cite{irboyrai,irthres} for consistency of presentation,  and to set the stage for the study of the quantum master equation and the dynamics of thermalization. We consider a model of bosonic fields $\Phi;\chi_{1,2}$ with a trilinear coupling described by the Lagrangian density

\be
\mathcal{L} = \frac{1}{2}\,\partial^\mu \Phi  \partial_\mu \Phi - \frac{1}{2}\,M^2 \Phi^2 +
\frac{1}{2}\,\partial^\mu \chi_1\,  \partial_\mu \chi_1- \frac{1}{2} \,m^2_1 \,\chi^2_1 +\frac{1}{2}\,\partial^\mu \chi_2\,  \partial_\mu \chi_2- \frac{1}{2} \, m^2_2 \,\chi^2_2\,-
\lambda \,\Phi     \chi_1 \chi_2 \,, \label{lagsuper}\ee yielding the total Hamiltonian
\be H= H_0 +H_I \label{htota}\ee with $H_0 = H_0[\phi]+H_0[\chi] $ the free field part and $H_I$ the interaction. The interaction vertex and conventions for the fields are depicted in fig.(\ref{fig:vertex}), in the following we will refer collectively as $\chi \equiv \chi_1,\chi_2$.

By adjusting the various masses, we can study the cases that yield threshold and infrared divergences within the same model allowing us to extract more general conclusions.

\begin{figure}[ht!]
\begin{center}
\includegraphics[height=2.5in,width=2.5in,keepaspectratio=true]{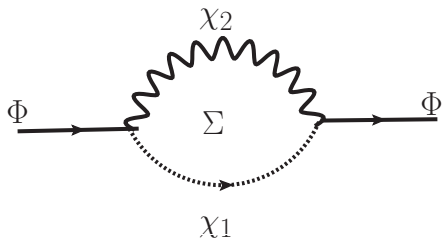}
\caption{The   one loop self energy $\Sigma(P^2)$  of the $\Phi$ field.}
\label{fig:selfenergy}
\end{center}
\end{figure}

The spectral properties of the $\Phi$ particle are summarized in the Dyson-resummed propagator including self energy corrections, the one loop self energy is shown in fig.(\ref{fig:selfenergy}).  A study of the Kallen-Lehmann representation of the $\Phi$ propagator has been presented in ref.\cite{irthres} revealing threshold singularities when $M$ coincides with the two particle threshold $M= m_1+m_2$ and infrared singularities when $m_1=M, m_2=0$. We summarize the main aspects of these divergences in order to establish a clear relation to the situation wherein the particles $\chi_{1,2}$ are in a thermal bath and the relaxation of $\Phi$ is studied via a quantum master equation.

\textbf{Threshold singularity.} The case when $M$ coincides with the two particle threshold $M = m_1+m_2$ has been studied  in ref.\cite{will} and more recently in  ref.\cite{irthres},   we consider  the   case $m_1=m_2=m$, which provides simpler expressions.  The ultraviolet divergence of the self energy is absorbed into mass renormalization and we redefine  the renormalized mass as $M$. For $M < 2m$ the Dyson resummed propagator (with the one loop self energy) features an isolated single particle pole below the two particle threshold, with residue (wave function renormalization)
\be   Z  = \Bigg[ 1-\frac{\partial \, {\Sigma}(P^2)}{\partial P^2}\Bigg]^{-1}_{P^2=M^2}\,, \label{Zdef} \ee where $\Sigma(P^2)$ is the $\Phi$ self energy.  As discussed previously in ref.\cite{will} and more recently in ref.\cite{irthres} $\partial \Sigma(P^2)/\partial P^2|_{P^2=M^2}$ features a singularity as $M\rightarrow 2 m$ from below, this is the origin  of the threshold singularity, with the consequence that $Z$ vanishes as $M^2\rightarrow 4 m^2$ as
\be Z  ~~~ {}_{ \overrightarrow{M\rightarrow 2m}} ~~~
       \frac{\sqrt{\frac{4m^2}{M^2}-1}}{\pi\,g^2}\,, \label{vanishzeta}\ee where
       \be g = \frac{\lambda}{4\pi\,M}\,,  \label{coupling} \ee is the effective dimensionless coupling. As the mass shell merges with the two particle threshold, the single particle state is no longer an asymptotic state, however the $\Phi$ particle does not decay as in the usual exponential decay case $e^{-\Gamma t}$ because the S-matrix decay rate
       \be \Gamma_q = \pi\,g^2\,\frac{M^2}{\Omega_q} \,\sqrt{1-\frac{4m^2}{M^2}}\,, \label{gammapole}\ee vanishes at $M^2 = 4m^2$.   A dynamical resummation method in real time implemented in ref.\cite{irthres} revealed that the survival probability of an initial single particle state $\ket{1^{\Phi}_{\vq}}$ of momentum $\vq$, and energy $\Omega_q = \sqrt{q^2+M^2}$,   is asymptotically given by
       \be |\langle 1^{\Phi}_{\vq}|U(t,0)| 1^{\Phi}_{\vq}\rangle|^2 = e^{-\sqrt{t/t^*_q}}~~;~~ t^*_q = \frac{\Omega_q}{4\pi\,g^4\,M^2}\,,  \label{survprob} \ee where $U(t,0) = e^{iH_0 t}\,e^{-i H t}$ is the unitary time evolution operator in the interaction picture, with $H_0$ the free field   and $H$ the full interacting Hamiltonians respectively. The square root behavior of the survival probability is a consequence of fact that the spectral density vanishes as a square root near threshold\cite{irthres} and this, in turn, is a consequence of the threshold singularity manifest in the vanishing of the single particle residue as $M$ approaches the threshold $2m$, the $\Phi$ particle is no longer an asymptotic state. The vanishing of the residue $Z$ as $M\rightarrow 2m$ is, therefore, a harbinger of the ``decay'' of the $\Phi$ particle, even when the S-matrix decay rate vanishes at $M=2m$.   As discussed in detail in ref.\cite{irthres} the decay is a consequence of off-shell processes of small virtuality $\propto 1/t$ at long time.

\vspace{1mm}

\textbf{Infrared singularity.} An infrared singularity in    $\partial\Sigma(P^2)/\partial P^2|_{P^2=M^2}$ arises when $m_1=M$ and $m_2=0$, this situation corresponds to the emission and absorption of massless quanta by a massive particle and is similar to the infrared divergence in the electron propagator in Quantum Electrodynamics\cite{bn,lee,chung,kino,kibble,yennie,weinberg,kulish,greco}. In order to understand the emergence of the infrared singularity more clearly, let us consider that $m_1=m;m_2=0$ and explore the limit $M \rightarrow m$ from below in which the infrared divergence becomes manifest. For $M<m$ the Dyson-resummed propagator including the one loop self energy features a single particle pole below the two particle continuum beginning at $P^2 = m^2$. However, as shown in ref.\cite{irthres} as $M\rightarrow m$ from below $\partial \Sigma(P^2)/\partial P^2|_{P^2=M^2}$ features an infrared singularity and as a consequence of this infrared divergence the  residue at the single particle pole vanishes, namely
\be Z ~~{}_{\overrightarrow{M\rightarrow m}} ~~ \frac{1}{g^2 \,\ln\big[ \frac{m}{m-M}\big]} \,.\label{irZ} \ee The vanishing of the residue entails that the $\Phi$ particle is not an asymptotic state, however it does not decay in the usual exponential manner because the S-matrix decay rate
\be \Gamma = \pi\, g^2\, \frac{M^2}{\Omega_q} \Big[ 1- \frac{m^2}{M^2} \Big] \,, \label{rateir} \ee vanishes for $M=m$. The dynamical resummation method introduced in refs.\cite{irboyrai,irthres} reveals that the survival probability decays asymptotically as a power law with anomalous dimension
\be |\langle 1^{\Phi}_{\vq}|U(t,0)| 1^{\Phi}_{\vq}\rangle|^2 = \Big[\Omega_q t\Big]^{-2g^2}  \,,  \label{survprobir} \ee again as a consequence of off-shell processes of small virtuality $\propto 1/t$ at long time.

Therefore, when $\partial \Sigma(P^2) /\partial P^2|_{P^2 = M^2}$ features either a threshold or infrared divergence, the amplitude of an initial single particle state decays in time, not as an exponential but with a decay law described above in each case, even when the \emph{on shell} decay rates vanish. In both cases the decay is not described by an on-shell process, with energy momentum conservation because the phase space for decay   calculated within the S-matrix framework vanishes in both cases, but by off-shell processes of small virtuality $\propto 1/t$ in the long time limit.

The question that we now address is, if and how, $\Phi$ particles thermalize when they couple to a heat bath  of particles $\chi_{1,2}$ in thermal equilibrium, when the respective masses yield threshold and infrared divergences.

 \section{The quantum Master equation:}\label{sec:master}

 In most approaches to quantum kinetics either collisional kernels that input S-matrix, on-shell transition rates, or alternative formulations that input on-shell spectral densities are invoked. Instead, we seek a formulation that just like the dynamical resummation method of refs.\cite{irboyrai,irthres} describes the time evolution and relaxation in terms of time dependent rates, thereby allowing virtual processes when the field $\Phi$ is coupled to a thermal bath in equilibrium of the fields $\chi_1,\chi_2$ in the cases of threshold and infrared divergences.

 Motivated by the theory of open quantum systems ubiquitous in quantum optics and quantum information\cite{breuer,zoeller}, we adapt the Lindblad formulation of the quantum master equation for the reduced density matrix of  the field $\Phi$ to include off-shell processes.

 The quantum master equation in a Lindblad form\cite{lin,gori,pearle,weinberg1,weinberg2,weinberg3}   has   recently received attention in applications to high energy physics\cite{banks,openburra,openaka,openyao,openmiura,openbram,boyopen}. This formulation begins with the time evolution of an initial density matrix that describes the total system of fields $\Phi,\chi_{1,2}$, which is given by
  \be \hat{\rho}(t) = e^{-iHt}\hat{\rho}(0)e^{iHt}\,, \label{rhooft} \ee with $H$ the total Hamiltonian. In the master equation approach\cite{breuer,zoeller} the time evolution of the density matrix is considered in the interaction picture. With  the full density matrix $\hat{\rho}(t)$  given by eqn. (\ref{rhooft}) the density matrix in the interaction picture is given by
\be \hat{\rho}_I(t)= e^{iH_0 t} \hat{\rho}(t) e^{-iH_0t}\,,\label{rhoIP}\ee whose time evolution obeys
\be \dot{\hat{\rho}}_I(t) = -i \big[H_I(t),\hat{\rho}_I(t)\big]\,, \label{rhodotip}\ee where $H_I(t)$ is the interaction Hamiltonian   in the interaction picture.   The solution of eqn. (\ref{rhodotip})  is given by
\be \hat{\rho}_I(t)= \hat{\rho}(0) -i \int^t_0 dt' \big[H_I(t'),\hat{\rho}_I(t')\big] \,. \label{solurhoip}\ee This solution is inserted back into (\ref{rhodotip}) leading to the iterative equation
\be \dot{\hat{\rho}}_I(t) = -i \big[H_I(t),\hat{\rho}(0)\big] -\int^t_0   \,\big[H_I(t),\big[H_I(t'),\hat{\rho}_I(t')\big]\big]\,dt' \,.\label{rhodotiter}\ee The next steps rely  on a series of \emph{assumptions}\cite{breuer}:

\textbf{\underline{i): Factorization}}: the total density matrix factorizes into a direct product of the density matrix for the $\Phi$ field, $\hat{\rho}_{I\Phi}(t)$ and that of the bath of $\chi$ fields, $\hat{\rho}_\chi$, namely,
\be \hat{\rho}_I(t) = \hat{\rho}_{I\Phi}(t)\otimes \hat{\rho}_\chi(0)\,, \label{factorization}\ee under the \emph{assumption}  that the   bath degrees of freedom   remain in thermal equilibrium, hence the density matrix of the bath does not depend on time. The \emph{reduced density matrix} for the field $\Phi$ is obtained by taking the trace of the full density matrix over the bath degrees of freedom, which by assumption remains in thermal equilibrium, therefore
\be  \hat{\rho}_{I\Phi}(t)  = \mathrm{Tr}_{\chi} \hat{\rho}_I(t) \,. \label{redmtxfi}\ee

\begin{figure}[ht!]
\begin{center}
\includegraphics[height=2.5in,width=2.5in,keepaspectratio=true]{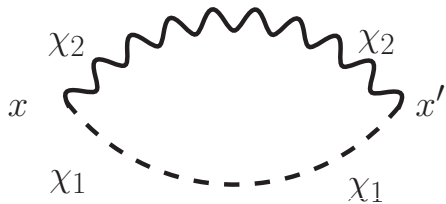}
\caption{Correlation functions $G^>(x-x'),G^<(x-x')$.}
\label{fig:correlator}
\end{center}
\end{figure}

 Upon taking the trace over the $\chi_{1,2}$ degrees of freedom  the first term on the right hand side of eqn. (\ref{rhodotiter}) vanishes, and we find the evolution equation for the  reduced density matrix for $\Phi$ in the interaction picture,
\bea \dot{\hat{\rho}}_{I\Phi}(t)   & =  & -\lambda^2\int^t_0 dt' \int d^3x \int d^3 x'\Bigg\{ \Phi(x) \, \Phi(x')\,\hat{\rho}_{I\Phi}(t') \,\,G^>(x-x') + \hat{\rho}_{I\Phi}(t') \,\Phi(x')\, \Phi(x)\,G^<(x-x') \nonumber \\
& - & \Phi(x) \,\hat{\rho}_{I\Phi}(t') \, \Phi(x')\, G^<(x-x') - \Phi(x') \, \hat{\rho}_{I\Phi}(t')\, \Phi(x)\,G^>(x-x') \Bigg\} \label{Linblad}\eea
where   we use the shorthand convention $x \equiv (\vec{x},t)~;~x' \equiv (\vec{x}',t')$, and introduced the bath correlation functions
\bea  G^>(x-x') & = &  \mathrm{Tr}_{\chi} \hat{\rho}_\chi(0) \chi_1(x)\chi_2(x) \chi_1(x')\chi_2(x') \label{ggreat} \\
G^<(x-x') & = &  \mathrm{Tr}_{\chi} \hat{\rho}_\chi(0) \chi_1(x')\chi_2(x') \chi_1(x)\chi_2(x)\,.  \label{gless} \eea
These correlation functions are displayed in fig.(\ref{fig:correlator}) and are directly related to the self-energy of the $\Phi$ field shown in fig.(\ref{fig:selfenergy}).

  At this stage a second approximation is  invoked:

   \underline{\textbf{ii:) Markov approximation}} which entails replacing $\rho_{I\Phi}(t') \rightarrow \rho_{I\Phi}(t)$ in the time integral. This approximation is  justified in weak coupling, as can be seen by   considering the first term in (\ref{Linblad}) as an example, it can be written as
\be -\lambda^2 \Phi(\vx,t) \int^t_0 \frac{d\mathcal{J}(t')}{dt'} \,\hat{\rho}_{I\Phi}(t') \,dt' ~~;~~  \mathcal{J}(t') \equiv \int^{t'}_0 \hat{\Phi}({\vx}',t'') \, G^>(\vx-{\vx}',t-t'')dt'' \label{incha}\ee which upon integration by parts yields
\be -\lambda^2 \Phi(\vx,t)  \mathcal{J}(t) \hat{\rho}_{I\Phi}(t) + \lambda^2  \Phi(\vx,t) \int^t_0  \mathcal{J}(t') \,\frac{d\hat{\rho}_{I\Phi}(t')}{dt'} dt' \label{incha2}\ee in the second term $d\hat{\rho}_{I\Phi}(t')/dt' \propto \lambda^2$ so this   term yields a contribution that is formally of order $\lambda^4$ and can be neglected to second order. The same analysis can be applied to all the other terms in (\ref{Linblad}) with the conclusion that in weak coupling and  to leading order $(\lambda^2)$  the Markovian approximation  $\hat{\rho}_{I\Phi}(t') \rightarrow \hat{\rho}_{I\Phi}(t)$ is justified.

Therefore in the Markov approximation the quantum master equation becomes
\bea \dot{\hat{\rho}}_{I\phi}(t)   & =  & -\lambda^2\int^t_0 dt' \int d^3x \int d^3 x'\Bigg\{ \Phi(x) \, \Phi(x')\,\hat{\rho}_{I\Phi}(t) \,\,G^>(x-x') + \hat{\rho}_{I\Phi}(t) \,\Phi(x')\, \Phi(x)\,G^<(x-x') \nonumber \\
& - & \Phi(x) \,\hat{\rho}_{I\Phi}(t) \, \Phi(x')\, G^<(x-x') - \Phi(x') \, \hat{\rho}_{I\Phi}(t)\, \Phi(x)\,G^>(x-x') \Bigg\} \,. \label{markovlim}\eea However, in contrast with the usual approach\cite{breuer,zoeller} to the Lindblad form of the quantum master equation, we \emph{will not} take the infinite time limit in the upper limit of the integral in $t'$ in eqn. (\ref{markovlim}), keeping the upper limit $t$ finite. This is a noteworthy difference with most previous approaches to the quantum master equation, which leads to time dependent rates and ultimately to allowing the off-shell processes with small virtuality to play a fundamental role in the dynamics of relaxation in the cases with threshold and infrared divergences.

The correlation functions $G^{>}(x-x'),G^<(x-x')$ are obtained in appendix (\ref{app:correlations}) in terms of spectral representations. They are given by
\bea  G^>(x-x') & = & \frac{1}{V}\sum_{\vq} \int dq_0 ~\varrho^>(q_0,\vq)\,e^{-iq_0(t-t')}~e^{i \vq\cdot(\vx-\vx')} \label{ggreatspec}\\
G^<(x-x') & = & \frac{1}{V}\sum_{\vq} \int dq_0 ~\varrho^<(q_0,\vq)\,e^{-iq_0(t-t')}~e^{i \vq\cdot(\vx-\vx')} \,,\label{glessspec}\eea where the spectral densities obey the relation
\be \varrho^>(-q_0,\vq)= \varrho^<(q_0,\vq)\,,\label{rela3}  \ee and fulfill the Kubo-Martin-Schwinger condition\cite{kms}
\be \varrho^<(q_0,\vq) = e^{-\beta\,q_0} \,\varrho^>(q_0,\vq)\,, \label{kms} \ee
which  is  a consequence of the fact that the fields $\chi_1,\chi_2$ are in thermal equilibrium. Introducing the spectral density
\be \varrho(q_0, \vq) = \varrho^>(q_0,\vq)-\varrho^<(q_0,\vq)\,,  \label{specdens}\ee the Kubo-Martin-Schwinger condition (\ref{kms}) leads to the following relations
\bea \varrho^>(q_0,\vq) & = &  [1+n(q_0)]\,\varrho(q_0,\vq) \label{rhogreat2} \\\varrho^<(q_0,\vq) & = &   n(q_0)\,\varrho(q_0,\vq) \label{rholess2} \eea where $n(q_0) = [e^{\beta\,q_0}-1]^{-1}$ is the Bose-Einstein distribution function at temperature $T=1/\beta$.  The above relations are proven in appendix (\ref{app:correlations}).

The spectral density $\varrho(q_0,\vq)$ is obtained in appendices (\ref{app:specdensthre},\ref{app:specdensir}) for the cases  of threshold and infrared divergences respectively.

In the interaction picture the fields feature the free field time evolution, therefore, upon quantization in a finite volume $V$ (eventually taken to infinity), we expand the field in the interaction picture as

\be \Phi(\vx,t)=  \sum_{\vq} \frac{1}{\sqrt{2\,V\Omega_q}}\Big[a_{\vq}\,e^{-i\Omega_q t} + a^\dagger_{-\vq} \, e^{i\Omega_q t} \Big]\,   e^{i\vq\cdot\vx}\, \label{aadag}\ee   where the annihilation and creation  operators $a_{\vq};a^\dagger_{-\vq}$ do not depend on time. At this point we invoke yet another approximation\cite{breuer,zoeller}:

 \underline{\textbf{iii:) ``rotating wave approximation'':}}  in writing the products $\Phi(\vx,t)~\Phi(\vx',t')$ of interaction picture field operators (\ref{aadag}) in (\ref{Linblad}) there are two types of terms  with very different time evolution. Terms of the form
\be a^\dagger_{\vq}~a_{\vq}~e^{i{\Omega_q(t-t')}}\,, \label{ada} \ee and its hermitian conjugate are ``slow'', and terms of the form
\be a^\dagger_{\vq}~a^\dagger_{-\vq}~ e^{2i\Omega_q t}~e^{i{\Omega_q(t-t')}}~~;~~ a_{\vq}\,a_{-\vq}~ e^{-2i\Omega_q t}\,e^{-i{\Omega_q(t-t')}}\,, \label{adad}\ee are fast, the extra rapidly varying phases $e^{ \pm 2i\Omega_q t}$ lead to rapid dephasing and do not yield resonant (nearly energy conserving) contributions. These terms only give perturbatively small transient contributions and are discussed in section (\ref{sec:discussion}). Keeping only the slow terms which dominate the long time dynamics  for $t \gg 1/\Omega_q$ and neglecting the fast oscillatory terms defines the ``rotating wave approximation'' ubiquitous in quantum optics\cite{breuer,zoeller}.

We will adopt these approximations and comment in section (\ref{sec:discussion}) on the corrections associated with keeping the fast   terms as well as caveats in the factorization approximation.

%%%% new addition

It is worth emphasizing that our approach, while following most of the same steps as those leading to the usual Lindblad quantum master equation\cite{breuer}, \emph{implies one less approximation}: whereas in the usual approach the infinite time limit is taken in the integrals defining the rates, we keep the finite time limits. In this sense, the approach advocated in this study adopts less approximations than the usual one.

%%% end of new addition.

Implementing the \textbf{Markov}  approximation $\hat{\rho}_{I\Phi}(t')\rightarrow \hat{\rho}_{I\Phi}(t)$, and the   \textbf{rotating wave } approximation (keeping only terms of the form $a^\dagger~a, a~a^\dagger$) using the spectral representation of the correlators (\ref{ggreatspec},\ref{glessspec}) with the property $\rho^<(-q_0,\vq)=\rho^>(q_0,\vq)$  and carrying out the spatial and temporal integrals we obtain the  \emph{Lindblad} form\cite{breuer,zoeller,lin,gori,pearle,weinberg1,weinberg2,weinberg3} of the quantum master equation, but with \emph{time dependent rates}, namely
\bea \dot{\hat{\rho}}_{I\phi}(t)  & = &  \sum_{\vk} \Bigg\{   -i\,\Delta_k(t)~\Big[a^\dagger_{\vk}\,a_{\vk}, \hat{\rho}_{I\phi}(t) \Big] \nonumber \\
& - & \frac{\Gamma^>_k(t)}{2} \Big[a^\dagger_{\vk}\,a_{\vk}~ \hat{\rho}_{I\phi}(t) + \hat{\rho}_{I\phi}(t)~a^\dagger_{\vk}\,a_{\vk} - 2 a_{\vk}~\hat{\rho}_{I\phi}(t)~a^\dagger_{\vk} \Big]\nonumber \\
& - & \frac{\Gamma^<_k(t)}{2} \Big[a_{\vk}\,a^\dagger_{\vk}~ \hat{\rho}_{I\phi}(t) + \hat{\rho}_{I\phi}(t)~a_{\vk}\,a^\dagger_{\vk} - 2 a^\dagger_{\vk}~\hat{\rho}_{I\phi}(t)~a_{\vk} \Big] \Bigg\} \,,
\label{Linfin} \eea  where
\be \Delta_k(t) = \frac{\lambda^2}{2\,\Ok} \,\int dk_0 \,\varrho(k_0,k)\,\frac{\Big[1-\cos[(\Ok-k_0)t] \Big]}{(\Ok-k_0)}\,, \label{Roftim}\ee
\be \Gamma^>_k(t) = \frac{\lambda^2}{\Ok}  \,\int dk_0 \, \varrho(k_0,k)\,\big[1+n(k_0)\big]\frac{ \sin[(\Ok-k_0)t] }{ (\Ok-k_0)} \,, \label{gamgre}\ee
\be \Gamma^<_k(t) = \frac{\lambda^2}{\Ok}  \,\int dk_0 \, \varrho(k_0,k)\, n(k_0) \frac{ \sin[(\Ok-k_0)t] }{(\Ok-k_0)} \,, \label{gamles}\ee
and we introduce
\be \Gamma_k(t) = \Gamma^>_k(t)-\Gamma^<_k(t)= \frac{\lambda^2}{\Ok} \,\int dk_0 \,\varrho(k_0,k)\,\frac{ \sin[(\Ok-k_0)t] }{(\Ok-k_0)}\,. \label{gamadif}\ee

The second and third lines in (\ref{Linfin}) are called the \emph{dissipator}, these are non-Hamiltonian, purely dissipative terms. In refs. \cite{lin,gori,pearle,weinberg2} it is argued that the equation (\ref{Linfin}) is the most general linear evolution equation that preserves unit trace and  Hermiticity of the density matrix.

 If at this stage  we take  the formal long time limit and replace $\sin[(\Ok-k_0)t] /(\Ok-k_0) \rightarrow \pi \delta(\Ok-k_0)$ as is usual in the derivation of Fermi's golden rule, we would obtain
\be \Gamma^>_k(t) ~~ {}_{\overrightarrow{t\rightarrow \infty}} ~~   \frac{\pi\, \lambda^2}{\Ok} \, \varrho(\Ok,k)\,\big[1+n(\Ok)\big] \equiv \Gamma^>_k  = \big[1+n(\Ok)\big]\,\Gamma_k \,,\label{gamaglontim}\ee
\be \Gamma^<_k(t) ~~ {}_{\overrightarrow{t\rightarrow \infty}} ~~   \frac{\pi\, \lambda^2}{\Ok} \, \varrho(\Ok,k)\, n(\Ok)  \equiv \Gamma^<_k =  n(\Ok)\,\,\Gamma_k \,,\label{gamallontim}\ee
 where
 \be \Gamma^>_k(t)-\Gamma^<_k(t) ~~{}_{\overrightarrow{t\rightarrow \infty}} ~~ \equiv \Gamma_k  =   \frac{\pi\, \lambda^2}{\Ok} \, \rho(\Ok,k) \,,   \label{gamadiflontim}\ee
 is \emph{precisely} the ``on-shell'' rate obtained from Fermi's Golden rule. However, in the cases under consideration describing threshold and infrared instabilities, the zero temperature ``on-shell'' rate vanishes  (see eqns. (\ref{rhothres}),(\ref{rhoir1}) below).

As explicitly shown below, keeping the time finite allows off-shell processes with small virtuality $\propto 1/t$ at  long but finite time, which will be ultimately responsible for thermalization in the cases of threshold and infrared divergences in which the on-shell rates vanish.

For any interaction picture operator $\mathcal{O}$ associated with the   field $\Phi$
\be \frac{d}{dt}\langle \mathcal{O}\rangle = \mathrm{Tr}_{\Phi}\Big\{ \dot{\mathcal{O}}~\hat{\rho}_{I\Phi}(t) + {\mathcal{O}}~\dot{\hat{\rho}}_{I\Phi}(t)\Big\} \,, \label{timederave}\ee where the average $\langle (\cdots) \rangle = \mathrm{Tr}_{\Phi}(\cdots)\hat{\rho}_{I\Phi}(t)$. Because $a_{\vk} ,a^\dagger_{\vk}$ are time independent in the interaction picture, the expectation value of the number operator
\be N_q(t) = \mathrm{Tr}_{\Phi}\, \hat{\rho}_{I\Phi}(t)\, a^\dagger_{\vq} \, a_{\vq}\,\label{numop}\ee obeys the quantum kinetic equation

\be \frac{dN_q(t)}{dt} =  \mathrm{Tr}_{\Phi}\Big\{ a^\dagger_{\vq} \, a_{\vq} ~\dot{\hat{\rho}}_{I\Phi}(t)\Big\} = -\Gamma_q(t) N_q(t) + \Gamma^<_q(t) \,. \label{dNdt}\ee

Similarly, we also find the evolution equation for the averages
\bea \frac{d}{dt}\langle a_{\vk} \rangle(t)  & = &  \Big[-i\,\Delta_k(t)-\frac{\Gamma_k(t)}{2} \Big]\, \langle a_{\vk} \rangle (t)\nonumber \\\frac{d}{dt}\langle a^\dagger_{\vk}\rangle(t)  & = &  \Big[i\,\Delta_k(t)-\frac{\Gamma_k(t)}{2} \Big]\, \langle a^\dagger_{\vk} \rangle (t)\,, \label{aveaad}\eea
and for the off-diagonal coherences,
\bea \frac{d}{dt} \langle a_{\vk}~ a_{-\vk} \rangle (t) & = &  \Big[-2i\,\Delta_k(t)  - \Gamma_k(t)\Big] \langle a_{\vk}~ a_{-\vk} \rangle (t) \nonumber \\
 \frac{d}{dt} \langle a^\dagger_{\vk}~ a^\dagger_{-\vk} \rangle(t)  & = &  \Big[2i\,\Delta_k(t)  - \Gamma_k(t)\Big] \langle a^\dagger_{\vk} ~ a^\dagger_{-\vk} \rangle(t)\, .  \label{bilins}\eea From the evolution equations (\ref{aveaad},\ref{bilins}) it is clear that $\Delta_k(t)$ is a time dependent renormalization of the frequency $\Omega_k$. Assuming that the initial averages $\langle a_{\vk} \rangle(0)=0~;~\langle a_{\vk}~ a_{-\vk} \rangle (0)=0$ such values remain as fixed points of the evolution equations.

 If the rates $\Gamma^{\lessgtr}(t)$ remain finite in the infinite time limit, replacing them in (\ref{dNdt}) by the formal long time limits (\ref{gamaglontim}-\ref{gamadiflontim})  as in Fermi's Golden rule,  the rate equation (\ref{dNdt}) would yield the solution
\be N_q(t) = n(\Omega_q)+\big[N_q(0)-n(\Omega_q)\big]\,e^{-\Gamma_q t}~~;~~  n(\Omega_q)=\frac{1}{e^{\beta \,\Omega_q}-1}\label{thermalasy} \ee which describes thermalization, and an exponential approach to the thermal fixed point of the quantum kinetic equation.   However, as discussed in detail below for the threshold and infrared singular cases under consideration, the rate $\Gamma_q=0$, which would imply that the distribution function does not evolve in time. However, as we show in detail below, in these cases keeping the time dependence in $\Gamma^{<}(t),\Gamma^>(t)$ the distribution function does evolve in time and ultimately reaches thermal equilibrium with the bath in a manner that cannot be described within Fermi's golden rule, or by extracting the rates from S-matrix theory. Taking   the long time limit in the rates  prior to solving the full quantum kinetic equation imposes strict energy conservation thereby neglecting off shell processes with small virtuality $\propto 1/t$ but with important consequences.

The full solution of the rate equation (\ref{dNdt}) is given by
\be N_q(t)=e^{-\gamma(t)}~\Big[N_q(0)+\int_0^t \Gamma^<_q(t')~e^{\gamma(t')} \,dt' \Big] \,, \label{noftnosec}\ee with
\be \gamma(t) \equiv \int^t_0 \Gamma_q(t')dt' = 2 \int^\infty_{-\infty} \tilde{\varrho}(q_0,q)\, \frac{\Big[1-\cos\big[\big(\Omega_q-q_0 \big)\,t \big] \Big]}{\big(\Omega_q-q_0 \big)^2}~~dq_0  \,, \label{integama} \ee
\be \Gamma^<_q(t) =     2\,\int^\infty_{-\infty}   \tilde{\varrho}(q_0,q)\, n(q_0) \frac{ \sin[(\Omega_q-q_0)t] }{(\Omega_q-q_0)}\,dq_0 \,, \label{gamles2}\ee where we defined
\be \tilde{\varrho}(q_0,q) = \frac{\lambda^2}{2\,\Omega_q}\, {\varrho}(q_0,q)\,.\label{tilvarrho}\ee

And the solutions of eqns. (\ref{aveaad},\ref{bilins}) are, respectively
\be \langle a_{\vk} \rangle(t)  = e^{-i \delta \Omega_k(t)\,t}\, e^{-\frac{\gamma(t)}{2} }\,\langle a_{\vk} \rangle(0)\,, \label{soluat}\ee
\be \langle a_{\vk}~ a_{-\vk} \rangle(t)= e^{-2 i \delta \Omega_k(t)\,t}\, e^{-\gamma(t)}\,\langle  a_{\vk}~ a_{-\vk} \rangle(0)\,,\label{solbilt} \ee and their hermitian conjugates. In the long time limit
\be \delta \Omega_k(t) =   \int^\infty_{-\infty}   \frac{\tilde{\varrho}(q_0,q)}{\Omega_k-q_0} \,\Big[1- \frac{\sin[(\Omega_k-q_0)t]}{(\Omega_k-q_0)t} \Big] \,dq_0 ~~{}_{\overrightarrow{t\rightarrow \infty}}~~\delta \Omega_k(\infty) = \int^\infty_{-\infty}  \mathcal{P}\Bigg[ \frac{\tilde{\varrho}(q_0,q)}{(\Omega_k-q_0)} \Bigg]\,dq_0\,\label{renfreq}\ee is a renormalization of the frequency $\Omega_k$ and $\mathcal{P}$ stands for the principal part.

\vspace{2mm}

\section{Threshold and infrared singular cases:}\label{sec:cases}
Armed with the general results (\ref{noftnosec},\ref{integama},\ref{soluat},\ref{solbilt}) we can now address the cases that feature threshold and infrared divergences, for which we need the corresponding spectral densities.

Since the spectral density is an odd function of $q_0$, and in order to more clearly highlight the regions of support of the time dependent functions,  it is convenient to implement the results of appendix (\ref{app:correlations})  and write

\be \int^t_0 \Gamma_q(t')dt' = 2 \int^\infty_{-\infty} \big[\tilde{\varrho}^{I}(q_0,q)+\tilde{\varrho}^{II}(q_0,q)\big] \,
\big[C_-(q_0,t)-C_+(q_0,t)\big] \,dq_0 \,, \label{intgam2} \ee
\be \Gamma^<_q(t) =     2\,\int^\infty_{-\infty}    \big[\tilde{\varrho}^{I}(q_0,q)+\tilde{\varrho}^{II}(q_0,q)\big]\big[ S_-(q_0,t)\,n(q_0) - S_+(q_0,t)\,n(-q_0) \big] \,dq_0 \,, \label{gamles22}\ee

where

\be C_{\mp}(q_0,t) = \frac{\Big[1-\cos\big[\big(\Omega_q \mp q_0 \big)\,t \big] \Big]}{\big(\Omega_q\mp q_0 \big)^2} ~~;~~ S_{\mp}(q_0,t) = \frac{\sin\big[\big(\Omega_q \mp q_0 \big)\,t ]}{(\Omega_q \mp q_0 )} \,, \label{cosins} \ee and $\tilde{\varrho}^{I}(q_0,q),\tilde{\varrho}^{II}(q_0,q)$ are given by equations (\ref{varro1},\ref{varro2})  and are obtained for the cases of threshold and infrared singularites in appendices (\ref{app:specdensthre},\ref{app:specdensir}) respectively.

At long time the functions $C_{\mp}(q_0,t)$ and $S_{\mp}(q_0,t)$ are strongly peaked at   $q_0 = \pm \Omega_q$, within a region of width $\simeq 2\pi/t$. Therefore the long time behavior of the rates (\ref{intgam2},\ref{gamles22}) are determined by the regions of the spectral density with support near $q_0 = \pm \Omega_q$, we refer to these as the resonance regions.

\subsection{Threshold singularity:} To exhibit the threshold singularity in its simplest manifestation, we consider that both $\chi_1,\chi_2$ fields have the same mass, namely $m_1=m_2=m$ and the two particle threshold coincides with the ``mass shell'' of the $\Phi$ particle, namely $M^2=4m^2$. From the expressions (\ref{intgam2},\ref{gamles22}) the long time dynamics is dominated by the regions of the spectral density $q_0 \simeq  \pm \Omega_q$, in other words, for $q^2_0 \simeq  \Omega^2_q = q^2 + M^2 $, near the mass shell of the $\Phi$ particle which in this case coincides with the two particle cut at $4m^2 = M^2$.

 As found in appendix (\ref{app:specdensthre}) $\tilde{\varrho}^{I}(q_0,q)$ only has support in the resonance region $q_0 \simeq \Omega_q$ for $q_0 >0$, and $\tilde{\varrho}^{II}(q_0,q)$ only features support below the light cone $q^2_0 < q^2$  far away from the resonance region, see equations (\ref{2parts},\ref{ld}). The oscillatory contributions from the non-resonant terms average out   or yield   a perturbatively small constant in the long time limit which can be safely neglected. Therefore in this case we can neglect in (\ref{intgam2},\ref{gamles22}) the contributions from $\tilde{\varrho}^{II}$ because it features support far away from $q_0 \simeq \pm \Omega_q$, and $C_+(q_0,t);S_+(q_o,t)$  because these are non-resonant in the region of support of the spectral density. Hence, in this case
 \be \gamma(t) \equiv \int^t_0 \Gamma_q(t')dt'     =  2 \int^\infty_{-\infty}  \tilde{\varrho}^{I}(q_0,q)  \,
 \frac{\Big[1-\cos\big[\big(\Omega_q - q_0 \big)\,t   \Big]}{\big(\Omega_q- q_0 \big)^2} \, \,dq_0 \,, \label{intgam3} \ee
 \be \Gamma^<_q(t) =     2\,\int^\infty_{-\infty}     \tilde{\varrho}^{I}(q_0,q)\,   \frac{\sin\big[\big(\Omega_q - q_0 \big)\,t ]}{(\Omega_q - q_0 )} \, n(q_0)   \,dq_0 \,, \label{gamles23}\ee where (see eqn. (\ref{2parts}) with $M^2=4m^2$)
 \be \tilde{\varrho}^{I}(q_0,q) = \frac{\lambda^2}{32\,\pi^2  \,\Omega_q} \, \Bigg\{ \Bigg[ \frac{q^2_0 - \Omega^2_q}{q^2_0 - q^2}\Bigg]^{1/2} + \frac{2\,T}{q} \, \ln\Big[\frac{1-e^{-\beta E^+}}{1-e^{-\beta E^-}} \Big] \Bigg\}\Theta(q_0 - \Omega_q) \,,\label{rhothres}\ee with
 \be E^{\pm} = \frac{\Omega_q}{2}+ \varepsilon^{\pm}~~;~~ \varepsilon^{\pm} = \frac{1}{2} \Bigg\{(q_0-\Omega_q) \pm q\,\Bigg[ \frac{q^2_0 - \Omega^2_q}{q^2_0 - q^2}\Bigg]^{1/2} \Bigg\} \,,  \label{epmth}     \ee where we have separated the terms $\varepsilon^{\pm} $ in the
 expressions for $E^{\pm}$ since these terms vanish as $q_0\rightarrow \Omega_q$ which is the dominant region of the spectral density in the long time limit.

 Notice that because $\varepsilon^\pm$ vanish  at threshold $q_0 = \Omega_q$, it follows that  $\tilde{\varrho}^I(\Omega_q,q) =0$, hence taking the infinite time limit in the rates, leading to    Fermi's golden rule (\ref{gamadiflontim}) would result in $\Gamma^> = \Gamma^< = 0$ and no equilibration. However, this is a consequence of taking the infinite time limit too soon, thereby neglecting processes with small virtuality $\propto 1/t$, as the analysis below shows in detail.

 Let us first focus on the integral defining $\gamma(t)$ in eqn. (\ref{intgam3}). Introduce the following dimensionless quantities
 \be \eta = \frac{q_0-\Omega_q}{\Omega_q} ~~;~~ \tau = \Omega_q \,t \,,\label{dimles} \ee in terms of which
 \be \gamma(\tau) =   \int^\infty_0 \sigma(\eta,q)\, \frac{1-\cos(\eta \tau)}{\eta^2}\,d\eta ~~;~~ \sigma(\eta,q)= \frac{2\,\tilde{\varrho}^{I}(q_0,q)}{\Omega_q}\big|_{q_0 = \Omega_q(1+\eta)} \,. \label{nuint}\ee In the long time limit $\tau \rightarrow \infty$ the function $(1-\cos(\eta\tau))/\eta^2$ is strongly peaked at $\eta \simeq 0$ with a height $\propto \tau^2 $ and  very narrow width $\propto 1/\tau$, therefore   it is the region near threshold, namely $q_0 \simeq \Omega_q$ or $\eta \simeq 0$ that dominates $\gamma(t)$ in the long time limit. We separate this region by writing
 \be \gamma(\tau) = \gamma_1(\tau) + \gamma_2(\tau)\,, \label{gamasplit}\ee where
\be \gamma_1(\tau) = \int^1_0 \sigma(\eta,q)\, \frac{1-\cos(\eta \tau)}{\eta^2}\,d\eta ~~;~~\gamma_2(\tau) = \int^\infty_1 \sigma(\eta,q)\, \frac{1-\cos(\eta \tau)}{\eta^2}\,d\eta \,, \label{gamase} \ee whereas in $\gamma_1(\tau)$ we must keep the $1/\eta^2$ together with the $-\cos(\eta\tau)/\eta^2$ because of the singularity at $\eta =0$, in $\gamma_2$ there is no such singularity in the domain of integration and we can separate these terms, the $-\cos(\eta\tau)/\eta^2$ oscillates rapidly and averages out in the $\tau \rightarrow \infty$ limit (Riemann-Lebesgue lemma) and we conclude that
\be \gamma_2(\tau)_{~~\overrightarrow{\tau \rightarrow \infty}} \int^\infty_1   \frac{\sigma(\eta,q)}{\eta^2} \,d\eta\, ,\label{lotigama2} \ee namely a time independent constant. In $\gamma_1$ it is now convenient to change variables to
\be \eta = \frac{x}{\tau}\,, \label{nux}\ee yielding
\be \gamma_1(\tau) = \tau\,\int^\tau_0 \sigma\big(\frac{x}{\tau},q\big)\, \frac{1-\cos(x)}{x^2}\,dx \,,  \label{rescagama1}\ee  this representation makes explicit that for $\tau \gg 1$  it is the region $x/\tau \ll 1$ that dominates, because the region $x\simeq \tau$ yields a contribution $\mathcal{O}(1/\tau) \ll 1$ to $\gamma_1$. Therefore, retracing the definition of variables, this analysis confirms that the long time limit of $\gamma(t)$ is dominated by the threshold region in the $q_0$ integrals, namely $q_0 -\Omega_q \propto 1/t$.

\begin{figure}[ht!]
\begin{center}
\includegraphics[height=3.5in,width=3.5in,keepaspectratio=true]{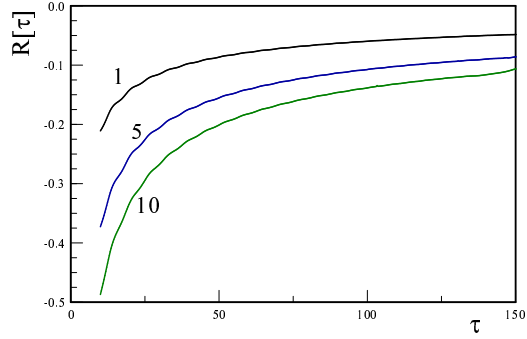}
\caption{The ratio  $R[\tau]$  (\ref{ratioap}) for $v=0.5$, and $\beta \Omega_q =1,5,10$ respectively.}
\label{fig:ratio}
\end{center}
\end{figure}

We note that $\varepsilon^{\pm}$ in eqn. (\ref{epmth}) vanish as $q_0 \rightarrow \Omega_q$, therefore in the long time limit, we can expand the exponentials $e^{-\beta \Omega_q/2}\,e^{-\beta \varepsilon^{\pm}}$ in the logarithms in eqn. (\ref{rhothres}) in powers of $\varepsilon^{\pm}$ and keep the leading term, yielding
\be \ln\Big[\frac{1-e^{-\beta E^+}}{1-e^{-\beta E^-}} \Big] = \beta \,q\, n\big(\frac{\Omega_q}{2}\big)\,\Bigg[ \frac{q^2_0 - \Omega^2_q}{q^2_0 - q^2}\,\Bigg]^{1/2}+ \mathcal{O}((q_0-\Omega_q)^2)\,. \label{logasy1} \ee As shown below, in terms of the variable $\tau$ (\ref{dimles}) the expansion in $\varepsilon^{\pm}$ is valid for $\sqrt{\tau} \gg \beta \Omega_q$.
We confirm the validity of this expansion by numerically studying the finite temperature contribution to $\gamma_1(\tau)$ given by eqn. (\ref{rescagama1}) and comparing it to the asymptotic form (\ref{logasy1}). To this end we write the logarithm on the left hand side of (\ref{logasy1}) in terms of the variables $x,\tau$ defined by eqns. (\ref{dimles},\ref{nux}) above,  with
\be \beta E^{\pm}(x/\tau) = \frac{\beta \Omega_q}{2}\big(1+\frac{x}{\tau} \big)\pm \frac{1}{2}\,\Delta(x/\tau) ~~;~~ \Delta(x/\tau) = \beta \Omega_q\,v\,\sqrt{\frac{x}{\tau}}\,\Bigg[\frac{2+\frac{x}{\tau}}{\big(1+\frac{x}{\tau}\big)^2-v^2} \Bigg]^{\frac{1}{2}} \,,\label{epmvar} \ee where $v = q/\Omega_q$, obtaining
\be \mathcal{L}(x/\tau) \equiv \ln\Big[\frac{1-e^{-\beta E^+(x/\tau)}}{1-e^{-\beta E^-(x/\tau)}} \Big]  = \ln\Bigg[1+n(E^-)\,\big(1-e^{-\Delta(x/\tau)}\big) \Bigg]\,,\label{logas} \ee Up to an overall factor $2T/q$ this is the finite temperature contribution to $\sigma(x/\tau)$ that enters in the integral in eqn. (\ref{rescagama1}). As anticipated, the factor $\beta \Omega_q/\sqrt{\tau}$ in $\Delta(x/\tau)$ clearly shows that for $\sqrt{\tau} \gg \beta\Omega_q$ and $x \lesssim 1$ it follows that $\Delta(x/\tau) \ll 1$ and the logarithm can be approximated as in eqn. (\ref{logasy1}). In terms of the variables $x,\tau$, the approximate asymptotic form on the right hand side of (\ref{logasy1}) becomes
\be \mathcal{L}_{as}(x/\tau)\equiv \beta \,q\, n\big(\frac{\Omega_q}{2}\big)\,\Bigg[ \frac{q^2_0 - \Omega^2_q}{q^2_0 - q^2}\,\Bigg]^{1/2} = n\big(\frac{\Omega_q}{2}\big)\,\Delta(x/\tau)\,.\label{asyfor} \ee In order to assess the reliability of the approximation (\ref{logasy1}) in the integral in (\ref{rescagama1}) we introduce the ratio
\be R(\tau) = \frac{\int^\tau_0 \Big[\mathcal{L}(x/\tau)-\mathcal{L}_{as}(x/\tau) \Big]\big(\frac{1-\cos(x) }{x^2}\big)\,dx}{\int^\tau_0   \mathcal{L}_{as}(x/\tau)  \big(\frac{1-\cos(x) }{x^2}\big)\,dx} \,,\label{ratioap}\ee which is displayed in fig. (\ref{fig:ratio}) for $v=0.5$ and $\beta \Omega_q = 1,5,10$ respectively. Similar results are obtained for different values of $v$.

This figure confirms the analysis above and shows that the  approach to  the asymptotic form is delayed for larger $\beta \Omega_q$.

  Returning to the original variables, and replacing the finite temperature contribution by its asymptotic limit  to   the  leading order in $q_0-\Omega_q$, we find
 \be \tilde{\varrho}^{I}(q_0,q) = \frac{\lambda^2}{32\,\pi^2  \,\Omega_q} \,   \Bigg[ \frac{q^2_0 - \Omega^2_q}{q^2_0 - q^2}\,\Bigg]^{1/2}\,\Big[1+ 2\, n(\Omega_q/2) \Big] \,,\label{rho1nt}  \ee the term $n(\Omega_q/2)$ is a consequence of stimulated emission and absorption. This result leads to
 \be \sigma(\eta,q) = \sqrt{2}\,g^2 \,\frac{M}{\Omega_q} \Bigg[\frac{\eta\,(1+\frac{\eta}{2})}{1+ 2\,\frac{\Omega^2_q}{M^2}\,\eta+ \frac{\Omega^2_q}{M^2}\,\eta^2} \Bigg]^{1/2}  \,\Big[1+ 2\,n(\Omega_q/2) \Big] \,,\label{sigthre} \ee where we introduced the dimensionless coupling
 \be g =  \frac{\lambda}{4\pi\,M}  \,. \label{coup}\ee Hence, from eqns. (\ref{gamasplit}-\ref{rescagama1}) we find\footnote{We used the result $\int^\infty_0 (1-\cos(x))/x^{3/2} dx = \sqrt{2\pi}$.}   in the long time limit $t \gg 1/T \,, 1/\Omega_q$
\be \gamma(t) = \sqrt{\frac{t}{t^*_q}}\,\Big(1+ \mathcal{O}(1/\Omega_q\,t) + \cdots \Big)  \,,  \label{lontigm1} \ee where we introduced the relaxation time scale
\be t^*_q = \frac{1}{4\pi g^4} \frac{\Omega_q}{M^2}\, {\tanh^2\big[\frac{\beta\Omega_q}{4}\big]}\,. \label{trela} \ee This result reveals that at high temperature, $\beta \Omega_q \ll 1$ the relaxation time $t^*_q$ is dramatically shortened as compared to the zero temperature ($\beta \Omega_q \rightarrow \infty$) result as a consequence of stimulated emission and absorption. At low temperature the asymptotic form (\ref{logasy1}) for the finite temperature contribution takes longer to emerge, however, its contribution to the relaxation time is negligible, since it is mostly determined by the zero temperature contribution.

This analysis confirms that the long time dynamics is determined by the threshold behavior of the spectral density with virtuality $q_0 -\Omega_q \simeq 1/t$, therefore, in the expression for $\Gamma^<(t)$, eqn. (\ref{gamles23}) we can expand the distribution function
\be n(q_0) = n(\Omega_q) + (q_0-\Omega_q)\,\frac{d n(q_0)}{dq_0}\Big|_{q_0 = \Omega_q}+ \cdots \ee and the first term $n(\Omega_q)$ yields the dominant contribution in the long time limit, yielding in this limit

 \be \Gamma^<_q(t) =  n(\Omega_q)\,   2\,\int^\infty_{-\infty}     \tilde{\varrho}^{I}(q_0,q)\,   \frac{\sin\big[\big(\Omega_q - q_0 \big)\,t ]}{(\Omega_q - q_0 )}     \,dq_0 + \mathcal{O}(1/t) + \cdots = n(\Omega_q) \frac{d\gamma(t)}{dt} + \mathcal{O}(1/t) + \cdots \,, \label{gamles23lt}\ee This result has important consequences. Since $\gamma(t) \propto \sqrt{t}$ the long time limit of the integral in
 the solution of the rate equation (\ref{noftnosec}) is dominated by the region of integration near the upper limit, namely $t' \simeq t$ therefore we can implement  the expansion (\ref{gamles23lt}) for $\Gamma^<_q(t')$ inside the integral in (\ref{noftnosec}) yielding in this  limit
 \be N_q(t) = n(\Omega_q) + e^{-\sqrt{t/t^*_q}}\,N_q(0)\,,\label{Neqi} \ee namely the distribution function approaches the thermal equilibrium form. Furthermore, the expectation values (\ref{soluat},\ref{solbilt}) and their hermitian conjugates vanish asymptotically as
 \be \langle a_{\vk} \rangle(t)  = e^{-i \delta \Omega_k \,t}\, e^{-\frac{1}{2}\sqrt{t/t^*_q} }\,\langle a_{\vk} \rangle(0)\,, \label{soluateqi}\ee
\be \langle a_{\vk}~ a_{-\vk} \rangle(t)= e^{-2 i \delta \Omega_k\,t}\, e^{-\sqrt{t/t^*_q}}\,\langle  a_{\vk}~ a_{-\vk} \rangle(0)\,,\label{solbilteqi} \ee   hence the reduced density matrix describes an equilibrium state at temperature $1/\beta$ and is diagonal in the occupation number and momentum basis. This thermal fixed point is approached asymptotically exponentially as $e^{-\sqrt{t/t^*_q}}$ rather than the usual $e^{-\Gamma t}$ as is typically the case in situations when Fermi's Golden rule applies or the S-matrix rate on shell is non-vanishing. This is a distinct example of thermalization via off-shell processes: a long but finite time allows an energy uncertainty and off shell processes with ``virtuality'' $q_0 -\Omega_q \simeq 1/t$ which determine the relaxation towards equilibration.

\subsection{Infrared singularity:}
In this case $m_1=M;m_2=0$, and the results of appendix (\ref{app:specdensir}) show that the contributions to the spectral density that feature support near the resonance regions can be summarized as
\be \tilde{\varrho}(q_0,q) = \tilde{\varrho}^{I}(q_0,q)+\tilde{\varrho}^{II}_B(q_0,q)\,,\ee with
\be \tilde{\varrho}^{I}(q_0,\vq) = \frac{\lambda^2}{32\,\pi^2 \,\Omega_q} \, \Bigg\{ \Bigg[ \frac{q^2_0 - \Omega^2_q}{q^2_0 - q^2}\Bigg]  + \frac{T}{ q } \, \ln\Bigg[\Bigg(\frac{1-e^{-\beta E^+}}{1-e^{-\beta E^-}}\Bigg)\Bigg( \frac{1-e^{-\beta  \mathcal{E}^-}}{1-e^{- \beta \mathcal{E}^+ }} \Bigg)\Bigg]  \Bigg\}\Theta(q_0 - \Omega_q )\,,\label{rhoir1}\ee

\be \tilde{\varrho}^{II}_B(q_0,q)  =\frac{\lambda^2\,T}{32\,\pi^2  \,\Omega_q \, q} \, \left\{\ln\Bigg[\frac{1-e^{\beta\,\mathcal{E}^-}} {1-e^{\beta\,\mathcal{E}^+}} \Bigg]-\ln\Bigg[\frac{1-e^{-\beta\,E_{-}}} {1-e^{-\beta\,E_{+}}} \Bigg]\right\} \,\Theta(\Omega_q-q_0)\,\Theta(q_0-q)\,, \label{rhoir2}  \ee where
\be E^\pm =  q_0 - \frac{q^2_0 - \Omega^2_q}{2(q_0 \pm q)} ~~;~~ \mathcal{E}^{\pm} = \frac{q^2_0-\Omega^2_q}{2(q_0 \pm q)} \label{Epm}\,. \ee

Because the relevant part of the spectral density features support for $q_0 >0$, it follows that
 \be \gamma(t) \equiv \int^t_0 \Gamma_q(t')dt'     =  2 \int^\infty_{-\infty}  \Big[\tilde{\varrho}^{I}(q_0,q)+\tilde{\varrho}^{II}(q_0,q)\Big]  \,
 \frac{\Big[1-\cos\big[\big(\Omega_q - q_0 \big)\,t   \Big]}{\big(\Omega_q- q_0 \big)^2} \, \,dq_0  \equiv \gamma^{I}(t)+\gamma^{II}(t)\,, \label{intgamIR} \ee were we have separated the respective contributions from $\tilde{\varrho}^{I},\tilde{\varrho}^{II}$. We study each in turn   by separating the zero and finite temperature contributions to $\gamma^{I}(t)$. First, for $q_0 > \Omega_q$ we introduce the dimensionless variables (\ref{dimles})  in terms of which $\gamma^{I}(t)$ features the same form as in eqns. (\ref{nuint}) and the same arguments lead to a similar separation of the integral in the variable $\eta$ (see eqn. \ref{dimles}), namely $\gamma^{I}(\tau)=\gamma^{I}_1(\tau)+\gamma^{I}_2(\tau)$ as in eqns.(\ref{gamasplit},\ref{gamase}) where $\gamma^{I}_2(\tau)$ yields a perturbatively small constant contribution in the long time limit $\tau \rightarrow \infty$ which can be neglected.

  Let us first consider the zero temperature contribution to $ \gamma^{I}_1(\tau)$, denoted by $\gamma^{I}_{10}(\tau)$. Writing in terms of the dimensionless coupling  $g$ introduced in eqn. (\ref{coup})
 \be \sigma^{I}_0(\eta,q) =  \frac{\lambda^2}{16\pi^2\,\Omega^2_q}\Bigg[\frac{\eta\,(2+\eta) }{\frac{M^2}{\Omega^2_q}+2\eta+\eta^2}\Bigg]
  = 2\,g^2 \,\eta + g^2 \,\eta^2\,\Bigg[\frac{\frac{M^2}{\Omega^2_q}-4-2\eta }{\frac{M^2}{\Omega^2_q}+2\eta+\eta^2} \Bigg]\,,  \label{sigosplit} \ee we find
 \be \gamma^{I}_{10}(\tau) = 2g^2 \, \Big[\ln\big(\tau\,e^{\gamma_E}\big) - Ci[\tau]  \Big]+ g^2\, \int^1_0 \Bigg[\frac{\frac{M^2}{\Omega^2_q}-4-2\eta }{\frac{M^2}{\Omega^2_q}+2\eta+\eta^2} \Bigg]\,(1-\cos(\eta\,\tau))\,d\eta \label{gama1IRzeroT} \ee where $\gamma_E$ is Euler's constant and $Ci[\tau]$ is the cosine integral function which vanishes as $\tau \rightarrow \infty$. In this limit the $\cos(\eta \tau)$ in the second term in (\ref{gama1IRzeroT})   averages out and this contribution   approaches a perturbatively small constant. Hence we conclude that in the long time limit, the zero temperature contribution from $\tilde{\varrho}^{I}$ yields
 \be \gamma^{I}_{10}(t)~~{}_{\overrightarrow{\Omega_q\,t\rightarrow \infty}}~~ 2g^2 \,\Big[\ln[\Omega_q\,t]+ \mathrm{constant}\Big]\,,  \label{lontiIR1} \ee which is the result  obtained in refs.\cite{irboyrai,irthres}, and given above by eqn. (\ref{survprobir}).

 Let us now consider the finite temperature contribution, beginning with $\tilde{\varrho}^{I}$.   Although a detailed understanding of the time evolution would require a numerical integration in a large range of parameters, as argued above, and shown below explicitly, the long time limit is captured by the behavior of the spectral density near the resonance $q_0 \simeq \Omega_q$. We observe that with $E^\pm$ given by eqn. (\ref{Epm}) it follows that $E^+ = E^-$ for $q_0 = \Omega_q$, therefore the ratio of logarithms featuring $1-e^{-\beta E^{\pm}}$ in the finite temperature contribution in $\tilde{\varrho}^{I}$ (\ref{rhoir1}) vanishes in the Fermi's golden rule limit (\ref{gamaglontim}-\ref{gamadiflontim}). However as we show below, by considering the time evolution of the rates we find that these do indeed contribute in the long-time limit.

 As discussed above, the long time limit is determined by the region $q_0 - \Omega_q \simeq 1/t$, let us consider first the logarithmic contribution featuring $E^{\pm}$ in (\ref{rhoir1}). For $q_0 \simeq \Omega_q$, it follows from (\ref{Epm}) that $E^+ \simeq E^-$, therefore writing
 \be e^{-\beta E^+} = e^{-\beta E^-}\,e^{-\beta(E^+-E^-)} \simeq e^{-\beta E^-}\,\Big(1-\beta(E^+-E^-) + \cdots \Big) \ee to leading order in $q_0 - \Omega_q$ we find
 \be \frac{T}{ q } \, \ln \Bigg(\frac{1-e^{-\beta E^+}}{1-e^{-\beta E^-}}\Bigg) =  \Bigg[ \frac{q^2_0 - \Omega^2_q}{\Omega^2_q- q^2}\Bigg]\,n(\Omega_q) + \mathcal{O}\big((q_0-\Omega_q)^2\big)\,.\label{firstexp}   \ee  The first term on the right hand side
 combines with the zero temperature contribution and is interpreted as stimulated emission and absorption, yielding  a contribution $\propto (1+n(\Omega_q))\,\ln(\Omega_q t)$ to $\gamma^1(t)$ in the long time limit similar to the zero temperature one (\ref{lontiIR1}), whereas the second term ($\mathcal{O}\big((q_0-\Omega_q)^2\big)$) yields a contribution that falls off as $1/t$ in the long time limit.

 The term with the logarithms involving $\mathcal{E}^{\pm}$ in (\ref{rhoir1}) originate in the distribution function of the massless $\chi_2$ particles in the bath and  yield  an unexpected result because $\mathcal{E}^{\pm}\rightarrow 0$ as $q_0 \rightarrow \Omega_q$. Therefore, expanding $e^{-\beta \mathcal{E}^{\pm}} \simeq 1 -\beta\,\mathcal{E}^{\pm}+\frac{1}{2} \beta^2 (\mathcal{E}^{\pm})^2 +\cdots$ we find that this term yields
 \be \frac{T}{q} \ln\Bigg( \frac{1-e^{-\beta  \mathcal{E}^-}}{1-e^{- \beta \mathcal{E}^+ }} \Bigg) = \frac{T}{q} \ln\Big(\frac{1+v}{1-v}\Big) - \frac{2\,T}{\Omega_q}\, \frac{ \Omega^2_q}{ M^2}\,\Big(\frac{q_0-\Omega_q}{\Omega_q}\Big)  - \frac{1}{2} \Bigg[ \frac{q^2_0 - \Omega^2_q}{\Omega^2_q - q^2} \Bigg]+\mathcal{O}\big((q_0-\Omega_q)^2 \big)~~;~~v = \frac{q}{\Omega_q}\,.  \label{2ndtermir} \ee Combining this result with (\ref{firstexp}), we find  that near the resonance region the finite temperature contribution to $\tilde{\varrho}^{I}$ (\ref{rhoir1}) is  to leading order
 \be  \frac{T}{ q } \, \ln\Bigg[\Bigg(\frac{1-e^{-\beta E^+}}{1-e^{-\beta E^-}}\Bigg)\Bigg( \frac{1-e^{-\beta  \mathcal{E}^-}}{1-e^{- \beta \mathcal{E}^+ }} \Bigg)\Bigg] = \frac{T}{q} \ln\Big(\frac{1+v}{1-v}\Big) - \frac{2\,T}{\Omega_q}\, \frac{ \Omega^2_q}{ M^2}\,\Big(\frac{q_0-\Omega_q}{\Omega_q}\Big)  - \frac{1}{2} \Bigg[ \frac{q^2_0 - \Omega^2_q}{\Omega^2_q - q^2}\,\big(1-2n(\Omega_q)\big) \Bigg]\,.  \label{finTrho1ap} \ee
 This expansion is  valid  for $q_0 \simeq \Omega_q$, therefore, since the resonance region dominates at long time when $q_0-\Omega_q \simeq 1/ t$ and because of the $\beta$ in the exponentials, the expansion becomes valid  for $t \gg 1/T$.

 We confirm this analysis  numerically as follows: we write the right  hand side of (\ref{finTrho1ap})   in terms of $x/\tau\equiv (q_0-\Omega_q)/\Omega_q$, yielding
 \be \beta \mathcal{E}^{\pm}(x/\tau) = \frac{\beta \Omega_q}{ \tau}\, \frac{x\big(1+\frac{x}{2\tau}\big)}{1-\frac{x}{\tau}\pm v} ~~;~~ \beta E^{\pm}(x/\tau) = \beta \Omega_q \,\big(1+\frac{x}{\tau}\big)-\beta \mathcal{E}^{\pm}(x/\tau)\,, \label{dimvarsro1} \ee  and define

  \bea I[ {x}/{\tau}]  & = &  \ln\Bigg( \frac{1-e^{-\beta  \mathcal{E}^-(x/\tau)}}{1-e^{- \beta \mathcal{E}^+(x/\tau) }}\Bigg) + \, \ln \Bigg(\frac{1-e^{-\beta E^+(x/\tau)}}{1-e^{-\beta E^-(x/\tau)}}\Bigg)   \label{Ifu} \\
I_{as}[ {x}/{\tau}] & = &  \ln\Big(\frac{1+v}{1-v}\Big) -\frac{2\,v}{1-v^2}\frac{x}{\tau}-\beta \Omega_q\, v \,\frac{x}{\tau}\,\Bigg[ \frac{1+\frac{x}{2\tau}}{1-v^2}\Bigg]\,\Big(1-2n(\Omega_q) \Big)\,,\label{Ias} \eea where $I_{as}[ {x}/{\tau}]$ follows from the expansion (\ref{finTrho1ap}), and numerically
   evaluate the integrals
\be J[\tau] =\int^\tau_0  I[x/\tau] \,\frac{1-\cos(x)}{x^2}\,dx   ~~;~~ D[\tau] = \int^\tau_0 \Big(I[x/\tau]-I_{as}[x/\tau]\Big) \,\frac{1-\cos(x)}{x^2}\,dx  \,. \label{diffe} \ee Up to an overall constant and a factor $\tau$, the integral $J[\tau]$  corresponds to the finite temperature contribution   to $\tilde{\varrho}^{I}$, and   to eqn. (\ref{intgamIR}) after the change of variables $q_0 = \Omega_q(1+x/\tau)$, whereas $D[\tau]$ quantifies  the  difference with the asymptotic form (\ref{finTrho1ap}).

\begin{figure}[ht!]
\begin{center}
\includegraphics[height=3.5in,width=3.5in,keepaspectratio=true]{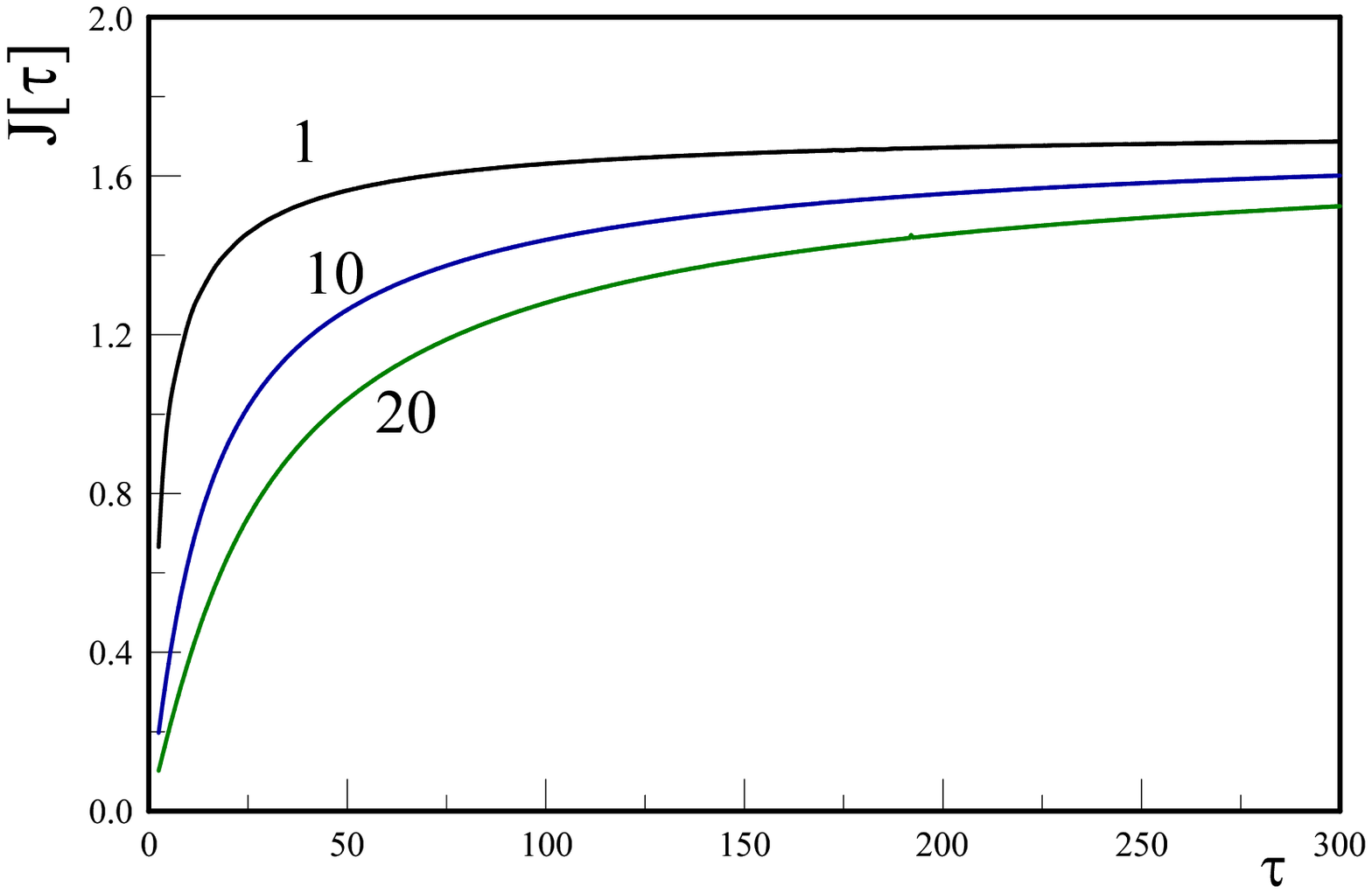}
\includegraphics[height=3.5in,width=3.5in,keepaspectratio=true]{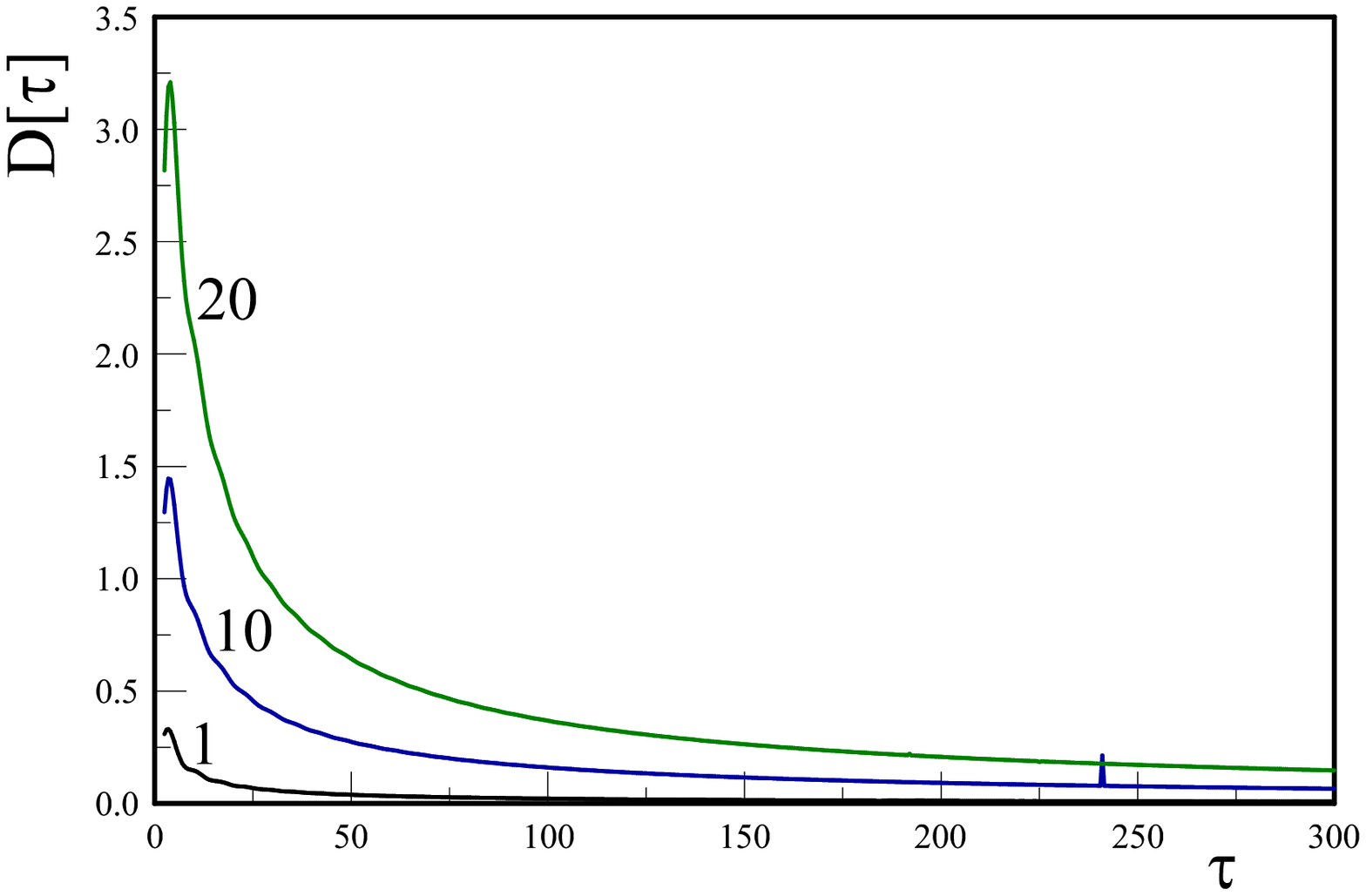}
\caption{The integrals $J[\tau]$ and $D[\tau]$, eqn. (\ref{diffe}) for $v=0.5$, and $\beta \Omega_q =1,10,20$ respectively.}
\label{fig:diffe}
\end{center}
\end{figure}

Fig. (\ref{fig:diffe}) shows these functions for $v=0.5$  highlighting the dependence of the time scales  on $\beta \Omega_q$, with  similar results for different values of $v$.

In particular the function $D[\tau]$, namely the difference between the exact and asymptotic form given by eqn.(\ref{2ndtermir}), shows the approach to the asymptotic behavior confirming the analysis above: the time scale of approach to the asymptotic limit increases with $\beta \Omega_q$, however at long time $\tau \gg \beta \Omega_q$ the approximation near threshold (\ref{2ndtermir}) reliably  describes the long time asymptotics.

Having confirmed quantitatively the validity of the analysis for the long time limit, we now combine the  results (\ref{firstexp}) and (\ref{2ndtermir}) with the zero temperature contribution to $\tilde{\varrho}^{I}$ and summarize its approximate form that describes the asymptotic long time limit, namely
\be \tilde{\varrho}^{I}_{as} (q_0,\vq) = \frac{\lambda^2}{32\,\pi^2 \,\Omega_q} \, \Bigg\{ \frac{1}{2} \Bigg[ \frac{q^2_0 - \Omega^2_q}{\Omega^2_q - q^2}\Bigg] \big(1+2 n(\Omega_q)\big) - \frac{2\,T}{\Omega_q}\, \frac{ \Omega^2_q}{ M^2}\,\Big(\frac{q_0-\Omega_q}{\Omega_q}\Big)  + \frac{T}{\Omega_q\,v}\, \ln\Big(\frac{1+v}{1-v}\Big) \Bigg\}\,\Theta(q_0-\Omega_q)\,.\label{varrho1asy}    \ee With this result we now provide an analytic expression that describes the asymptotic long time dynamics from  $\tilde{\varrho}^{I}$ implementing the following steps: i) we pass to the variables $\eta,\tau$ defined in eqn. (\ref{dimles}) and  introduce $\sigma(\eta,q)$ as per the definition in eqn. (\ref{nuint}), ii) we split the integral in the $\eta$ variable as in eqns. (\ref{gamasplit},\ref{gamase}), and neglect the contribution from $\gamma_2(\tau)$ which yields a perturbatively small constant in the long time limit, iii) we   split   the contribution of the first term in   (\ref{varrho1asy}) as in eqns. (\ref{sigosplit},\ref{gama1IRzeroT}) and find in the long time limit $t \gg 1/\Omega_q,1/T$,
 \be \gamma^{I}(\tau) = g^2 \,\Bigg\{\tau\,      \frac{\pi }{2\,\beta \Omega_q}\,  \frac{1-v^2}{v}  \ln\Big(\frac{1+v}{1-v}\Big) +  \ln(\tau)\,\Big[1+ 2n(\Omega_q)-\frac{2T}{\Omega_q} \Big] \Bigg\}\,. \label{gamaIfint}  \ee

In the heavy $\Phi$ particle limit with $v \ll 1$ this expression clearly shows a \emph{crossover} from a   $\ln(\tau)$ behavior at low temperature and $\tau   \ll  \beta \Omega_q \ln(\tau)$ to linear in time for $\tau \gg  \beta \Omega_q \ln(\tau)$, a stage that emerges at very long time at low temperature $\beta \Omega_q \gg 1$, or early on at very high temperature $\beta \Omega_q \ll 1$. However, as the $\Phi$ particle becomes ultrarelativistic $v\simeq 1$ this crossover occurs at a much later time even at large temperature, and the logarithmic growth in time dominates for  a much longer period.

 The contribution from $\gamma^{II}(t)$ must in principle be studied numerically, however, the lessons from the analysis carried out for $\gamma^I(t)$ can   now be implemented to obtain the asymptotic long time limit of $\gamma^{II}(t)$.

 The spectral density $\tilde{\varrho}^{II}_B$ (\ref{rhoir2}), is similar to the finite temperature contribution to  $\tilde{\varrho}^{I}$ (\ref{rhoir1})  but with a different domain $q \leq q_0 \leq \Omega_q$. The analysis above shows that the asymptotic long time limit    is obtained by expanding $\tilde{\varrho}^{II}_B$ near the resonance region $q_0 \simeq \Omega_q$, which in the case of $\tilde{\varrho}^{II}_B$ coincides with the \emph{upper threshold}. Performing the same approximations leading to the asymptotic form for $\tilde{\varrho}^{I}$, we find
 \bea \frac{ T}{  q} \, \left\{\ln\Bigg[\frac{1-e^{\beta\,\mathcal{E}^-}} {1-e^{\beta\,\mathcal{E}^+}} \Bigg]-\ln\Bigg[\frac{1-e^{-\beta\,E^{-}}} {1-e^{-\beta\,E^{+}}} \Bigg]\right\}  & =  & -\frac{1}{2}\Bigg[ \frac{\Omega^2_q-q^2_0}{\Omega^2_q-q^2}\Bigg]\,\big(1 + 2 n(\Omega_q)\big) + \frac{T}{q} \,\ln\Big[ \frac{1+v}{1-v}\Big] \nonumber \\ & + & \frac{2T\Omega_q }{M^2}\,\Big( \frac{\Omega_q-q_0}{\Omega_q}\Big)+ \mathcal{O}((\Omega_q-q_0)^2)\,,\label{rho2aSy}   \eea this expansion is valid near the upper threshold at $q_0 \simeq \Omega_q$, corresponding to the resonance region which dominates the long time dynamics. Because of the diffeent domain, we now introduce the dimensionless variable $y$ as $q_0 = \Omega_q (1-y/\tau)$,  in terms of which
 \be \beta \mathcal{E}^{\pm}(y/\tau) = -\frac{\beta \Omega_q}{ \tau}\, \frac{y\big(1-\frac{y}{2\tau}\big)}{1+\frac{y}{\tau}\pm v} ~~;~~ \beta E^{\pm}(y/\tau) = \beta \Omega_q \,\big(1-\frac{y}{\tau}\big)-\beta \mathcal{E}^{\pm}(y/\tau)\,, \label{dimvarsro2} \ee and define
  \bea  K[y/\tau]   &  =  &   \ln\Bigg[\frac{1-e^{\beta\,\mathcal{E}^-(y/\tau)}} {1-e^{\beta\,\mathcal{E}^+(y/\tau)}} \Bigg]-\ln\Bigg[\frac{1-e^{-\beta\,E^{-}(y/\tau)}} {1-e^{-\beta\,E^{+}(y/\tau)}} \Bigg] \nonumber \\ K_{as}[y/\tau]  & = & - \beta \Omega_q\,v\,\frac{y}{\tau}\frac{1-y/2\tau}{1 -v^2}\,\big(1 + 2 n(\Omega_q)\big)+ \ln \Big[ \frac{1+v}{1-v}\Big]+ \frac{2 v}{1-v^2}\,\frac{y}{\tau} \,, \label{ks} \eea we confirm the validity of the expansion (\ref{rho2aSy}) in the long time limit by studying numerically the integral
  \be H[\tau] = \int^{(1-v)\tau}_0 \Big(K[y/\tau]-K_{as}[y/\tau]\Big)\,\frac{1-\cos(y)}{y^2} \, dy \,, \label{Kint}\ee the upper limit reflects the lower threshold at $q_0 =q$ after the change of variables.

  Again, up to an overall constant and a factor $\tau$, the integral of $K[y/\tau]$  corresponds to the contribution from $\tilde{\varrho}^{II}$ to eqn. (\ref{intgamIR}) after the change of variables $q_0 = \Omega_q(1-y/\tau)$, in terms of which the upper threshold at $q_0 =\Omega_q$  is at $y=0$ and the lower threshold at $q_0 = q$ corresponds to $y = (1-v)\tau$. Note that in the ultrarelativistic limit $v \rightarrow 1$ the phase space for the contribution $\tilde{\varrho}^{II}$ vanishes.

\begin{figure}[ht!]
\begin{center}
\includegraphics[height=3.5in,width=3.5in,keepaspectratio=true]{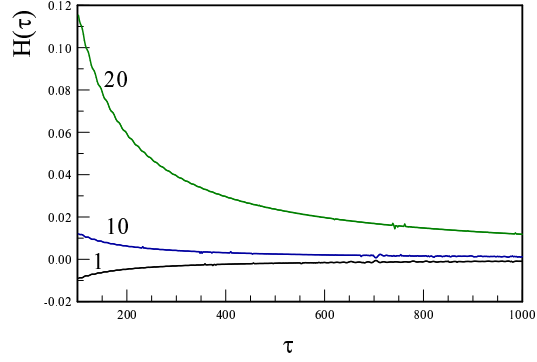}
\caption{The integral $H[\tau]$, eqn. (\ref{Kint}) for $v=0.5$, and $\beta \Omega_q =1,10,20$ respectively. The asymptotic behavior emerges at much later time for   $v \simeq 1$.}
\label{fig:diff2}
\end{center}
\end{figure}

   The function $H(\tau)$ describes the difference between the contribution with the full spectral density $\propto \tilde{\varrho}^{II}$ and the asymptotic form (\ref{rho2aSy}). It is displayed in fig. (\ref{fig:diff2}), showing the delayed approach to asymptotics: for large $\beta \Omega_q$ the asymptotic form is approached for $(1-v)\tau \gg \beta \Omega_q$. For values of $v \simeq 1$ the approach to the asymptotic behavior takes a much longer time, as a consequence of the closing off of the phase space for $\tilde{\varrho}^{II}$.

Therefore, the long time limit emerging for $t \gg T$ is described by
\be \tilde{\varrho}^{II}_{as}(q_0,q) = \frac{\lambda^2}{32\,\pi^2 \,\Omega_q} \,  \Bigg\{-\frac{1}{2}\Bigg[ \frac{\Omega^2_q-q^2_0}{\Omega^2_q-q^2}\Bigg]\,\big(1 + 2 n(\Omega_q)\big) + \frac{T}{\Omega_q\,v} \,\ln\Big[ \frac{1+v}{1-v}\Big]   +  \frac{2T}{\Omega_q}\frac{\Omega^2_q }{M^2}\,\Big( \frac{\Omega_q-q_0}{\Omega_q}\Big) \Bigg\}\,.\label{varo2asy} \ee Changing variables to $q_0 = \Omega_q(1-y/\tau)$ we   find
\bea \gamma^{II}(\tau) & = &  g^2 \Bigg\{ \Big[ Ci[(1-v)\tau]-\gamma_E-\ln[(1-v)\tau]\Big]\,\Big( 1 + 2 n(\Omega_q) - \frac{2T}{\Omega_q} \Big)\nonumber \\ & + & \frac{\tau\,T}{\Omega_q}\, \frac{1-v^2}{v}\,\ln\Big[\frac{1+v}{1-v} \Big] \Bigg[ Si[(1-v)\tau] - \frac{\Big(1-\cos[(1-v)\tau]\Big)}{(1-v)\tau}\Bigg]\Bigg\}\,. \label{gama2fina}  \eea Combining this result with $\gamma^{I}(\tau)$ eqn. (\ref{gamaIfint}) we finally find
\bea \gamma(\tau) & = &  g^2 \Bigg\{ \Big[ Ci[(1-v)\tau]-\gamma_E-\ln[(1-v)\tau]+\ln(\tau)\Big]\,\Big( 1 + 2 n(\Omega_q) - \frac{2}{\beta \Omega_q} \Big)\nonumber \\ & + & \frac{\tau }{\beta \Omega_q}\,  \frac{1-v^2}{v}\,\ln\Big[\frac{1+v}{1-v} \Big]\, \Bigg[\frac{\pi}{2}+ Si[(1-v)\tau] - \frac{\Big(1-\cos[(1-v)\tau]\Big)}{(1-v)\tau}\Bigg]\Bigg\}\,. \label{gamatotalir}  \eea This result exhibits several important features: \textbf{i:)} in the ultrarelativistic limit $v \simeq 1$ and for $\tau \ll 1/(1-v^2)$ the contribution $\gamma^{II}(\tau)$ given by eqn. (\ref{gama2fina}) is  negligibly small and $\gamma(\tau)$, is determined by $\gamma^{I}(\tau)$ which  features  a cross over from $\ln(\tau)$ to  linear in $\tau$ for $\tau \gg \beta \Omega_q\ln(\beta \Omega_q)  $. \textbf{ii):} For $v \ll 1$  and  $\tau \gg 1 $, $Ci[(1-v)\tau] \rightarrow 0~;~Si[(1-v)\tau] \rightarrow \pi/2$ and the logarithmic terms cancel each other, yielding a linear growth in time  $\gamma(t) \simeq 2 \pi\, g^2 \,t  \, T$. This behaviour can be summarized as
\bea \textbf{Case}~ 1 :  && v \simeq 1 ~~\mathrm{and}~~ \Omega_q\,t \ll 1/(1-v^2) \,\nonumber \\
&& \gamma(  t) \simeq g^2\,\Bigg\{  \begin{array}{c}
                                        \ln(\Omega_q\,t)~~\mathrm{for}~~ t \ll \beta \ln(\beta \Omega_q) \\
                                        \Omega_q \,t ~~\mathrm{for}~~ t \gg \beta \ln(\beta \Omega_q)
                                      \end{array} \,, \label{case1} \eea
\bea \textbf{Case}~ 2 :  && v \ll 1 ~~\mathrm{and}~~ \Omega_q\,t \gg 1  \,\nonumber \\
&& \gamma(  t) \simeq  2\pi g^2 T\,t\,. \label{case2} \eea

We now invoke the same argument as in the threshold case to show the emergence of thermalization. The integral in eqn. (\ref{noftnosec}) is dominated by the long time limit of $\gamma(t')$ which grows with time, and with $\Gamma^{<}_q(t)$ given by eqn. (\ref{gamles22}) only the term with $S_-(q_0,t)$ is resonant since the spectral density features support near the resonances only for $q_0 > 0$, and, as shown above the long time limit is dominated by the region near the resonance, $q_0 \simeq \Omega_q$. Therefore, in the long time limit we can expand
$n(q_0) = n(\Omega_q) + \mathcal{O}(q_0-\Omega_q)$ where the last term yields a contribution suppressed by $1/t$ in the long time limit. Therefore just as in the threshold case, but now with the contribution from $\tilde{\varrho}^{II}$ we find
 \be \Gamma^<_q(t) =  n(\Omega_q)\,   2\,\int^\infty_{-\infty}     \Big[\tilde{\varrho}^{I}(q_0,q)+\tilde{\varrho}^{II}(q_0,q) \Big]\,   \frac{\sin\big[\big(\Omega_q - q_0 \big)\,t ]}{(\Omega_q - q_0 )}     \,dq_0 + \mathcal{O}(1/t) + \cdots = n(\Omega_q) \frac{d\gamma(t)}{dt} + \mathcal{O}(1/t) + \cdots \,. \label{gamles23ltir}\ee Therefore, from eqn. (\ref{noftnosec}) we find that the asymptotic long time evolution of the occupation number is given by

 \be N_q(t) = n(\Omega_q) + e^{-\gamma(t)}\,N_q(0)\,,\label{NeqiIR} \ee namely the distribution function approaches the thermal equilibrium fixed point. Furthermore, just as in the threshold case the expectation values (\ref{soluat},\ref{solbilt}) and their hermitian conjugates vanish asymptotically as
 \be \langle a_{\vk} \rangle(t)  = e^{-i \delta \Omega_k \,t}\, e^{-\gamma(t)/2 }\,\langle a_{\vk} \rangle(0)\,, \label{soluateqiIR}\ee
 \be \langle a_{\vk}~ a_{-\vk} \rangle(t)= e^{-2 i \delta \Omega_k\,t}\, e^{-\gamma(t) }\,\langle  a_{\vk}~ a_{-\vk} \rangle(0)\,,\label{solbilteqiIR} \ee hence the reduced density matrix in this case also describes an equilibrium state at temperature $1/\beta$ diagonal in the occupation number and momentum basis. This thermal fixed point is approached with the function $\gamma(t)$ which now features various regimes with crossover between a logarithmic and a linear time dependence.

Having understood the long time limit, and the crossover between logarithmic and linear time dependence both in the non-relativistic and ultrarelativistic limits, we now focus on the following puzzling aspects of   $\tilde{\varrho}^{I},\tilde{\varrho}^{II}$,   and the final results for $\gamma^I(\tau)$ and $\gamma^{II}(\tau)$, these are:

\underline{\textbf{a:)}} What is the origin of the terms that survive in the zero temperature limit in eqns. (\ref{2ndtermir},\ref{rho2aSy})?.

\underline{\textbf{b:)}} What is the origin of the terms that \emph{do not vanish} as $q_0 \rightarrow \Omega_q$ in eqns. (\ref{2ndtermir},\ref{rho2aSy})?.

\underline{\textbf{c:)}} What is the origin of the cancellation of the logarithms between $\tilde{\varrho}^{I}$ and $\tilde{\varrho}^{II}$, in particular for $v \ll 1$?.

\vspace{1mm}

\textbf{The infrared at $T\neq 0$:} The answer to  \textbf{a)} and \textbf{b)}  originate in the integrals
 \be  \int^{E^+}_{E^-} n(q_0-E) dE \equiv \int^{\mathcal{E}^-}_{\mathcal{E}^+}n(\mathcal{E})\,d\mathcal{E}\,,\label{irintegs} \ee in eqns. (\ref{barrdosira},\ref{barrdosirb}), with $\mathcal{E}^{\pm}$ given by eqn. (\ref{Epm}), and $ n(q_0-E)$ corresponds to the distribution function of $\chi_2$.    As discussed above, in the long time limit it is the region $q_0 -\Omega_q \propto 1/t$ that dominates the $q_0$ integrals in the rates, and for $q_0 \simeq \Omega_q$ it follows from the expressions for $\mathcal{E}^{\pm}$ (\ref{Epm}),  that $\mathcal{E}^{\pm} \rightarrow 0$ as $q_0 \rightarrow \Omega_q$, hence the available phase space vanishes as the virtuality  $  q_0 -\Omega_q \propto 1/t$. The vanishing of $\mathcal{E}$ as $q_0 \rightarrow \Omega_q$ is a consequence of the masslessness of the $\chi_2$ particle, as shown in appendix (\ref{app:specdensir}) $\mathcal{E}$ is the momentum transferred to $\chi_2$, the massless field.  Therefore for $\mathcal{E}/T \ll 1$, or alternatively for small virtuality $q_0-\Omega_q \propto 1/t$, for $T t \gg 1$ we can expand
 \be n(\mathcal{E}) = \frac{T}{\mathcal{E}}-\frac{1}{2} +\mathcal{O} (\mathcal{E}/T)   \,, \label{expair}\ee which upon integration in (\ref{irintegs}) yields
 \be \int^{\mathcal{E}^-}_{\mathcal{E}^+}n(\mathcal{E})\,d\mathcal{E} = T \ln\Bigg[ \frac{q_0+q}{q_0-q}\Bigg]- \frac{q}{2} \Bigg[ \frac{q^2_0 - \Omega^2_q}{q^2_0 - q^2}\Bigg]+\cdots\,.\label{intirfini}   \ee Writing $q_0 = \Omega_q + (q_0-\Omega_q)$ and expanding  the logarithm  in $q_0-\Omega_q$ yields the terms  $\ln[(1+v)/(1-v)]$ in (\ref{varrho1asy},\ref{varo2asy}), the second term in   (\ref{intirfini}) yields  the temperature independent term which subtracts from the zero temperature contribution, thus  clarifying its origin in the limit $T t \gg 1$.  Because the virtuality $q_0 -\Omega_q \propto 1/t$ this limit corresponds to $T\gg \mathrm{virtuality}$. Therefore, although the phase space closes as $\mathcal{E}^{\pm} \rightarrow 0 $, the singular behavior of the distribution function for the massless $\chi_2$ for small virtuality  yields the finite temperature logarithmic correction and the temperature independent result. In turn, the finite temperature term $T\,\ln[(1+v)/(1-v)]$ yields the term that grows linearly with time in $\gamma(t)$. Hence, this linear time dependence is not a direct result of Fermi's golden rule, but a more subtle finite temperature infrared effect associated with the masslessness of $\chi_2$ in the   limit of small virtuality.

\vspace{1mm}

\textbf{Quantum kinetic interpretation of $\tilde{\varrho}^I,\tilde{\varrho}^{II}$}: The answer to \textbf{c)} is found in a simple, yet  illuminating interpretation of the processes that yield the spectral densities $\tilde{\varrho}^I,\tilde{\varrho}^{II}$, eqns. (\ref{varro1},\ref{varro2})  in terms of a linearized gain-loss master equation.

\begin{figure}[ht!]
\begin{center}
\includegraphics[height=3.5in,width=3.5in,keepaspectratio=true]{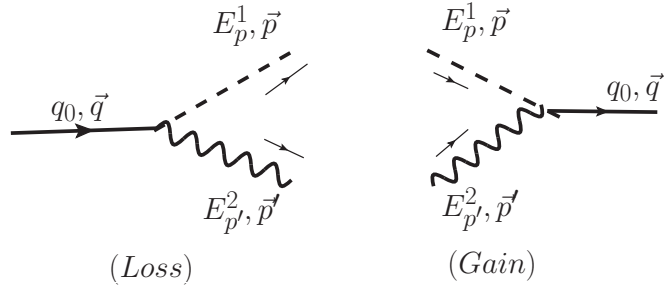}
\caption{The loss process $\Phi \rightarrow \chi_1 \chi_2$ and the inverse, gain process $\chi_1 \chi_2 \rightarrow \Phi$.}
\label{fig:qkone}
\end{center}
\end{figure}

Let us consider that the $\Phi$ particle features \emph{off-shell} energy $q_0$  and momentum  $\vec{q}$,  and an occupation $N_q(q_0,t)$. The usual quantum kinetic equation that implements Fermi's golden rule in the transition probabilities  is of the generic form gain-loss. Consider the loss  term from the ``decay'' $\Phi \rightarrow \chi_1 \chi_2$ and   the inverse, gain term,  from recombination $\chi_1 \chi_2 \rightarrow \Phi$  where $\chi_{1,2}$ are   in the bath in equilibrium, these processes are   shown in fig. (\ref{fig:qkone}). With the transition probabilities per unit time obtained as usual from S-matrix theory but considering the energy of the $\Phi$ particle as $q_0$ (allowing off-shellness), we find for these processes

\bea \frac{d N_q(q_0,t)}{dt}\Big|_{gain} & = & (1+N_q(q_0,t))\, \frac{\pi\,\lambda^2}{ \Omega_q} \int \frac{d^3p}{(2\pi)^3 2 E^1_p  2 E^2_{p'}} \,n(E^1_p)\,n(E^2_{p'})\, \delta(q_0-E^1_p -E^2_{p'})  \nonumber \\
\frac{d N_q(q_0,t)}{dt}\Big|_{loss} & = & N_q(q_0,t) \,\frac{\pi\,\lambda^2}{\Omega_q}  \int \frac{d^3p}{(2\pi)^3 2 E^1_p 2 E^2_{p'}} \,(1+n(E^1_p))\,(1+n(E^2_{p'}))\, \delta(q_0-E^1_p -E^2_{p'})     \nonumber \\ \vp' & = & \vq-\vp\,. \label{qkone}
 \eea

 It is straightforward to confirm that the equilibrium distribution $N_q(q_0) = n(q_0)$ is a fixed point of the gain-loss equation as a consequence of the energy   delta  functions and momentum conservation, therefore writing $N(q_0,t) = n(q_0)+\delta N(q_0,t) $ we find for the gain-loss equation
 \be \frac{d \,\delta N_q(q_0,t)}{dt} = - \delta N(q_0,t)\,2\pi\,\tilde{\varrho}^{I}(q_0,q)\,, \label{qkunovarro}\ee where $\tilde{\varrho}^I$ is given by eqn. (\ref{varro1}). This analysis clarifies that the origin of the $\tilde{\varrho}^I$ contribution are the gain and loss processes $\Phi \leftrightarrow \chi_1 \chi_2$  shown in fig. (\ref{fig:qkone}).

\begin{figure}[ht!]
\begin{center}
\includegraphics[height=3.5in,width=3.5in,keepaspectratio=true]{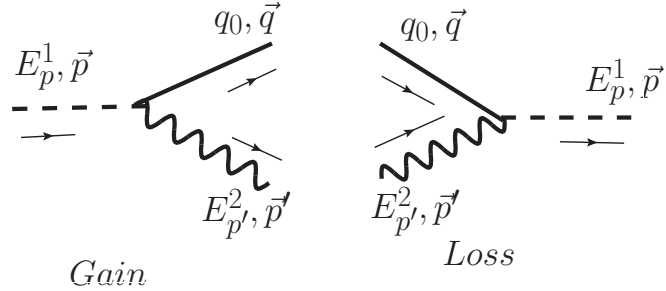}
\caption{The gain process $\chi_1   \rightarrow \Phi \chi_2$  and the inverse loss process $\chi_2 \,  \Phi \rightarrow  \chi_1$, where $\chi_{1,2}$ are in the thermal bath.}
\label{fig:qktwo}
\end{center}
\end{figure}

However, there are other gain and loss processes that also contribute, consider the gain process $\chi_1 \rightarrow \Phi\, \chi_2$    and the inverse loss process $\Phi\, \chi_2 \rightarrow \chi_1$, with $\chi_{1,2}$ being particles in the bath. These processes are shown in fig. (\ref{fig:qktwo}) and contribute to the gain and loss terms as

\bea \frac{d N_q(q_0,t)}{dt}\Big|_{gain} & = & (1+N_q(q_0,t))\, \frac{\pi\,\lambda^2}{ \Omega_q} \int \frac{d^3p}{(2\pi)^3 2 E^1_p  2 E^2_{p'}} \,n(E^1_p)\,(1+n(E^2_{p'}))\, \delta(E^1_p -q_0 -E^2_{p'})  \nonumber \\
\frac{d N_q(q_0,t)}{dt}\Big|_{loss} & = & N_q(q_0,t) \,\frac{\pi\,\lambda^2}{\Omega_q}  \int \frac{d^3p}{(2\pi)^3 2 E^1_p  2 E^2_{p'}} \,(1+n(E^1_p))\, n(E^2_{p'})\, \delta(E^1_p - q_0 -E^2_{p'})   \nonumber \\    \vp' & = & \vq-\vp \,.   \label{qktwo}
 \eea

Again, writing $N_q(q_0,t)= n(q_0) +\delta N_q(q_0,t)$ the equilibrium term vanishes and we find for the gain-loss linearized equation
 \be \frac{d \,\delta N_q(q_0,t)}{dt} = - \delta N(q_0,t)\,2\pi\,\tilde{\varrho}^{II}(q_0,q)\,, \label{qkdosvarro}\ee where $\tilde{\varrho}^{II}$ is given by eqn. (\ref{varro2}).   Exchanging $\chi_1 \rightarrow \chi_2$ is tantamount to replacing $\tilde{\varrho}^{II}(q_0,q) \rightarrow \tilde{\varrho}^{II}(-q_0,q)$.

 This analysis clarifies that the origin of the $\tilde{\varrho}^I$ contribution are the gain and loss processes $\Phi \leftrightarrow \chi_1 \chi_2$  shown in fig. (\ref{fig:qkone}) and those for $\tilde{\varrho}^{II}$ are the gain and loss processes $\chi_1 \leftrightarrow \Phi \chi_2$ displayed in fig.(\ref{fig:qktwo}).

 This interpretation also clarifies the mayor difference between the threshold and infared cases for $\tilde{\varrho}^{II}$. For the threshold case, with equal masses   $m_1=m_2=m$, and $M = 2m$  the kinematics for the processes $\Phi \leftrightarrow \chi_1\chi_2$ is very different from that of the processes $\chi_1 \leftrightarrow \Phi \chi_2$ because for $M = 2m$ the former occur very near the resonance at $q_0 \simeq \Omega_q$, whereas the latter is very far from it. Whereas in the infrared case with $m_1= M, m_2=0$ the process $\chi_1 \leftrightarrow \Phi \chi_2$ features the same kinematics as the processes $\Phi \rightarrow \chi_1 \chi_2$ since the mass of $\Phi$ coincides with that of $\chi_1$. This explains the similarity in the finite temperature contributions from $\tilde{\varrho}^{II}$ and
 $\tilde{\varrho}^{I}$ in the infrared case, as well as the cancellation of terms with $\ln(t)$ which arise from contributions of opposite signs between $\tilde{\varrho}^{I}$ and $\tilde{\varrho}^{II}$.

 It is important to highlight that we have considered an \emph{off shell} energy $q_0$ in this analysis, setting $q_0 =\Omega_q$ leads to vanishing contributions for $\tilde{\varrho}^{I}$ and $\tilde{\varrho}^{II}$ because of the kinematics of energy momentum conservation in both cases, threshold and infrared. Allowing the time evolution of the ``rates'' as in equations (\ref{integama},\ref{gamles2}) allows the uncertainty associated with a finite time interval to yield a non-vanishing time dependent rate which describe off shell processes of small virtuality $\propto 1/t$. Taking the infinite time limit from the outset as in the usual quantum kinetic approach implementing Fermi's golden rule with S-matrix transition rates imposes strict energy conservation thereby leading to the vanishing of the spectral functions on shell.

 \vspace{1mm}

\section{Discussion and caveats}\label{sec:discussion}

\textbf{Counterrotating terms:}

In the derivation of the  quantum master equation (\ref{Linfin}) we  neglected terms of the form
\be a_{\vk}\,a_{-\vk}\,e^{-2i\Ok t} e^{i\Ok(t-t')} ~~;~~ a^\dagger_{\vk}\,a^\dagger_{-\vk}\,e^{2i\Ok t} e^{-i\Ok(t-t')}\,. \label{counter}\ee The time integral over $t'$ can be carried out following the steps leading to equation (\ref{Linfin}) yielding contributions of the form
$a_{\vk}\,a_{-\vk}\,e^{-2i\Ok t} \rho^\lessgtr(k_0,k) \hat{\rho}_{I\Phi}(t)$ etc. The contribution of these terms to the equations of motion for linear or bilinear forms of $a,a^\dagger$ are straightforward to obtain, they do not yield terms that grow secularly in time because the rapid dephasing of the oscillatory terms average out in the time integrals. These are non-resonant terms and yield perturbatively small subleading contributions of the form $\delta \Ok/\Ok \ll 1~;~\Gamma_k/\Ok\ll 1$  in weak coupling, as compared to those obtained from equation (\ref{Linfin}) which captures the secular growth in time because of the resonances and describes the leading behavior in the long time dynamics.

\vspace{1mm}

\textbf{Quantum master equation and dynamical resummation:} The quantum master equation in Lindblad form obtained in this study, with
time dependent rates, provides a resummation of second order processes. The bath correlation functions (\ref{ggreat},\ref{gless}) displayed in fig.(\ref{fig:correlator})  are related to the self-energy of the $\Phi$ field in the thermal bath shown in fig.(\ref{fig:selfenergy}). Therefore, we conclude that the quantum master equation provides a resummation of one-particle irreducible diagrams in the bath and in real time, akin to the resummation in real time provided by the   method introduced in refs.\cite{irboyrai,irthres}. Just like the   latter framework yields a decaying survival probability as a consequence of processes with small virtuality in the long time limit, the master equation provides a similar time evolution but including medium effects that yield unexpected time dependence in the case of infrared singularity as a consequence of the infrared enhancement of the Bose  Einstein distribution function of massless particles.

\vspace{1mm}

\textbf{Factorization vs. entanglement:}  An important result of the study in refs.\cite{irboyrai,irthres} is that the asymptotic state is a kinematically entangled state of the ``daughter'' particles produced by the decay of the $\Phi$ field. This aspect is not shared by the master equation because of the assumption on factorization. This, one of the main assumptions, prevents the emergence of correlations between the $\chi_1,\chi_2$ density matrices, which is asummed to remain factorized and describing thermal equilibrium for each species. This is an important caveat and major difference with the dynamical resummation method which clearly shows entanglement in the final state. At this stage is not clear how to systematically include the correlations between the different components of the bath leading to entanglement and whether such correlation will influence the time evolution towards equilibration or possible new observable consequences. It is not a matter of keeping the higher orders in the coupling because the dynamical resummation method in refs.\cite{irboyrai,irthres} was implemented also up to second order in the coupling.   This aspect merits to be studied thoroughly in future work.

\vspace{1mm}

\textbf{Radiative corrections:} Radiative corrections may change the masses of the various fields. In the case of threshold divergences,
these may move the mass of the $\Phi$ field away from threshold, however if the corrections are perturbatively small (after renormalization), the results may apply nevertheless: if the corrections lead to a smaller $\Phi$ mass, but still perturbatively close to threshold small virtuality may still yield a relaxation towards equilibration, such is the case at zero temperature as discussed in ref.\cite{irthres}. If on the other hand finite temperature effects dramatically change the $\Phi$ mass then a new study with particular focus on the interplay between virtuality and distance to threshold will be required to assess whether virtual processes lead to equilibration. In the case of infrared divergences, unless the massless field is protected by a symmetry, such as a Goldstone field, radiative corrections will very likely induce a mass. If the mass term is perturbatively small, infrared enhancements will survive, if, on the other hand finite temperature effects yield a large radiatively induced mass, infrared effects may still arise for temperatures much larger than the mass, however, off-shell processes with small virtuality may be   important. All these possibilities would require  a reassessment and merit deeper understanding. Such program is clearly beyond the scope of this initial study.

\vspace{1mm}

\textbf{Higher order on-shell vs. off-shell small virtuality processes:} A competition between higher order on-shell processes and the lower order off-shell processes with small virtuality may emerge. As an example, consider the case of threshold divergences, which at lowest order yield exponential relaxation with exponent $\propto g^2\sqrt{t}$, and for example a scattering processes with on-shell in and out states yielding a time dependence $\propto g^4 t$. In the theory defined by the Lagrangian density (\ref{lagsuper}), for example the process $\Phi \Phi \rightarrow \chi_1 \chi_1$ with an intermediate $\chi_2$ propagator, which  at tree level yields a probability \emph{per unit time} $\propto g^4$, hence it contributes to the exponential in the survival probability with a term $\propto g^4 t$.  Such contribution becomes of the same order as the leading term $\propto g^2 \sqrt{t}$ at a time scale  $t  \propto  1/g^4$, namely $t \simeq t^*$, at which time the population and the expectation values have nearly reached their asymptotic values. Therefore, for weak coupling the leading order result captures the early, intermediate and long time asymptotics, and higher order corrections will quantitatively affect the very long time asymptotics. Obviously, these arguments will require a firmer quantitative assessment for particular theories with various different mass scales. In turn this entails obtaining a generalization of the Lindblad quantum master equation up to \emph{fourth order} in the interaction Hamiltonian, involving fourth order nested commutators with the density matrix. This generalization has not yet been explored in the literature, and is obviously well beyond the scope of this article.

\vspace{1mm}

\textbf{More general lessons:} Although this study has focused on the particular bosonic model described by the Lagrangian density (\ref{lagsuper}) the results allow to extract more general lessons. Above and beyond the  particular model, the main input in the Lindblad quantum master equation are the bath correlations in terms of the spectral densities of the bath degrees of freedom. As discussed previously the major difference in our study is that we do \emph{not} take the infinite time limit in the time integrals of the bath correlations, even when the Markov approximation remains one of the main assumptions. This feature is important because it allows the rates to depend explicitly on time thereby allowing virtual processes associated with the energy uncertainty in a finite time interval to play a fundamental role in the dynamics of relaxation. This is one of the main results of this study. These aspects transcend a particular model and suggest a more general range of applicability of the methods and results. For example in the case of a   $\Phi$ particle coupled to a bath with massless (gapless) degrees of freedom leading to infrared divergences, the vanishing residue of the quasiparticle and the crossover between $ln(t)$ and $\propto t$ behavior in the thermalization dynamics for $t\gg 1/T$ is reminiscent of the dynamics of heavy impurities in a Fermi sea and the orthogonality catastrophe\cite{orto,orto2}. Furthermore, recently\cite{levitov} it has been recognized that    photoexcitations of soft off-shell electron-hole pairs in graphene yield Sudakov (double logarithms) type of spectral densities with strong infrared behavior, therefore the results obtained in this study may prove relevant to study the possible thermalization of these excitations. We also expect that these methods may prove useful in cosmology, where the time dependence of the cosmological expansion may provide yet another route to virtual processes hitherto unexplored.

 \section{Conclusion and further questions:}\label{sec:conclusion}
 Motivated by   ubiquitous cross disciplinary interest, in this article we have   studied  the approach to thermalization via processes that cannot be described with  the usual quantum kinetic equations that input S-matrix, on-shell transition probabilities. These entail taking the infinite time limit thereby enforcing strict energy conservation and on-shell processes. Recent work\cite{irboyrai,irthres} which focused on the relaxation of single particle states in the case of threshold and infrared divergences at zero temperature highlighted the role of processes with small virtuality, which nonetheless lead to the ``decay'' of single particle states with unusual decay laws despite the    vanishing of on-shell decay rates.

 In this article we considered a model of a scalar fields $\Phi$ - the system- coupled to a bath   of scalar particles $\chi_{1,2}$ in thermal equilibrium, which allows to study the cases of threshold and infrared divergences by tuning the mass spectra.  Inspired by the theory of quantum open systems, we adapted a method  to study the relaxation of fields coupled to a thermal bath in equilibrium  in the cases when  the S-matrix transition rates vanish. We obtained a Lindblad type   quantum master equation for the reduced density matrix of the   $\Phi$ (system) field which departs from the usual form\cite{breuer,zoeller} by \emph{not} taking an infinite time limit, thereby allowing time dependent rates and processes with small virtuality. The resulting quantum master equation provides a \emph{real time} resummation of self-energy corrections determined by bath correlations at finite temperature. We find that in the case of threshold singularities, the reduced density matrix for $\Phi$   approaches a thermal fixed point as $e^{-\sqrt{t/t^*}}$ with the relaxation time $t^*$ shortening at high temperatures because of stimulated emission and absorption. In the case of infrared singularities, we find that a thermal fixed point is approached as  $e^{-\gamma(t)}$ with $\gamma(t)$ featuring a cross over from a $\propto \ln(t)$ to a $\propto t$ behavior for $t \gg 1/T$. This latter behavior is a result  of a subtle interplay between infrared enhancement at finite temperature and small virtuality. In both cases   the dynamics of relaxation and thermalization is dominated by off-shell processes with small virtuality $\propto 1/t$ in the long time limit.

 Although our study has focused on a particular model, the results are more overarching. The derivation of the quantum master equation maintaining finite time bath correlations input the spectral density of the bath correlations, therefore the modified Lindblad form can be adapted to other systems. Furthermore, the analysis highlights the virtues of small virtuality: processes that are forbidden by strict energy conservation, i.e. off-shell,  can nevertheless lead to thermalization with unusual dynamics of relaxation and a wealth of time scales towards equilibration that simply cannot be reliably captured with on-shell S-matrix transition probabilities. Furthermore, by allowing the rates in the quantum master equation to depend on time, it is possible that in some cases transient phenomena associated with small virtuality may actually compete with on-shell processes and contribute substantially to the dynamics of relaxation, this possibility merits further study.

 We also find noteworthy that a crossover from $\ln(t)$, to a linear $t$ behavior in the approach to thermalization in the case of infrared singularity, has also been found in the case of a heavy impurity in Fermi systems that feature an orthogonality catastrophe\cite{orto,orto2}. Furthermore,  off shell infrared phenomena   from photoexcitation of soft electron-hole pairs in graphene has been studied in ref.\cite{levitov}, and might possibly be yet another arena wherein the methods implemented here may prove useful.   An important area where we envisage possible applications is in cosmology where the universal expansion offers yet another route to hitherto unexplored virtual processes.

\acknowledgements
  The author gratefully acknowledges support from the U.S. National Science Foundation through grant award NSF 2111743.

\appendix
\section{Correlation functions:}\label{app:correlations}
The main ingredients to obtain the correlation functions (\ref{ggreat},\ref{gless}) are
\bea & & \mathrm{Tr}\rho_{\chi_a}(0)\,\chi_a(x)\chi_a(x')   =   \sum_{\vp} \frac{1}{2VE^{a}_{p}} \Big[(1+ n(E^a_p))\,e^{-iE^a_p (t-t')} + n(E^a_p)\,e^{iE^a_p (t-t')} \Big]\,e^{i\vp\cdot(\vx-\vx')} \nonumber \\
& & =  \sum_{\vp} \frac{1}{2VE^{a}_{p}} \,\int^\infty_{-\infty}    (1+ n(p_0)) \Big[\delta(p_0-E^a_p)-\delta(p_0+E^a_p)  \Big] \,e^{i\vp\cdot(\vx-\vx')}\,e^{-ip_0(t-t')}\,dp_0   \,, \label{correa} \eea  for $a= 1,2$, and where
\be n(p_0) = \frac{1}{e^{\beta p_0}-1} ~~;~~ n(-p_0) = - (1+n(p_0)) \,.\label{ocupa} \ee
From these expressions we obtain the dispersive representations (\ref{ggreatspec},\ref{glessspec}) with the spectral densities
\bea \varrho^>(q_0,\vq) &  =  & \frac{1}{V} \sum_{\vp^{\,\,'},\vp}\,\frac{\delta_{\vq,\vp+\vp^{\,\,'}}}{2\,E^{1}_{p}\,2\,E^{2}_{p^{\,'}}}\, \int dp_0 \,dp'_0 \,\delta(q_0-p_0-p'_0)(1+ n(p_0))(1+ n(p'_0)) \nonumber \\ & \times  & \Big[\delta(p_0-E^1_p)-\delta(p_0+E^1_p)  \Big] \Big[\delta(p'_0-E^2_{p'})-\delta(p'_0+E^2_{p'})  \Big] \label{barrogreat}\,,
 \eea

\bea \varrho^<(q_0,\vq) &  =  & \frac{1}{V} \sum_{\vp^{\,\,'},\vp}\,\frac{\delta_{\vq,\vp+\vp^{\,\,'}}}{2\,E^{1}_{p}\,2\,E^{2}_{p^{\,'}}}\,\int dp_0 \,dp'_0 \,\delta(q_0-p_0-p'_0) \, n(p_0)\,n(p'_0) \nonumber \\ & \times  & \Big[\delta(p_0-E^1_p)-\delta(p_0+E^1_p)  \Big] \Big[\delta(p'_0-E^2_{p'})-\delta(p'_0+E^2_{p'})  \Big] \label{barroless}\,.
 \eea

 Using the identity
 \be n(p_0) = e^{-\beta p_0}\,(1+n(p_0))\,,\label{ide}\ee and the $\delta(q_0-p_0-p'_0) $ we find the Kubo-Martin-Schwinger\cite{kms} relation
 \be \varrho^<(q_0,\vq) = e^{-\beta q_0}\,\varrho^>(q_0,\vq)\,, \label{kmsrel}\ee which is a direct consequence of the bath degrees of freedom being in thermal equilibrium.

 Defining
 \be \varrho(q_0,\vq) \equiv \varrho^>(q_0,\vq)-\varrho^<(q_0,\vq)\,,\label{specdif}\ee it follows from the relation (\ref{kmsrel}) that
 \be \varrho^>(q_0,\vq) = (1+n(q_0))\, \varrho(q_0,\vq) ~~;~~ \varrho^<(q_0,\vq) =  n(q_0)\, \varrho(q_0,\vq) \,. \label{rrhos}\ee Combining the results (\ref{barrogreat},\ref{barroless}) and taking the infinite volume limit, we find
 \bea \varrho(q_0,\vq)  & = & \int \frac{d^3p}{(2\pi)^3}\frac{1}{2\,E^{1}_{p}\,2\,E^{2}_{p^{\,'}}} \,\int dp_0 \,dp'_0 \,\delta(q_0-p_0-p'_0)(1+ n(p_0) + n(p'_0)) \nonumber \\ & \times  & \big[\delta(p_0-E^1_p)-\delta(p_0+E^1_p)  \big] \big[\delta(p'_0-E^2_{p'})-\delta(p'_0+E^2_{p'})  \big]~~;~~ \vp^{\,\,'} = \vq-\vp  \label{barrodif} \,. \eea

 %\be \tilde{\varrho}(q_0,q) = \frac{\lambda^2}{2\,\Omega_q}\, \Bigg\{ \Bigg\}  \ee

 Using the delta functions  and with the definition (\ref{tilvarrho}) we obtain the final form of the spectral density
 \be \tilde{\varrho}(q_0,q) = [\tilde{\varrho}^{I}(q_0,\vq)-\tilde{\varrho}^{I}(-q_0,\vq)]  + [\tilde{\varrho}^{II}(q_0,\vq)-\tilde{\varrho}^{II}(-q_0,\vq)]\,, \label{varrosplit} \ee where
 \be \tilde{\varrho}^{I}(q_0,\vq) = \frac{\lambda^2}{2\,\Omega_q}\, \int \frac{d^3p}{(2\pi)^3}\frac{1}{2\,E^{1}_{p}\,2\,E^{2}_{p^{\,'}}} \, \big[1+ n(E^{1}_{p}) +  n(E^{2}_{p^{\,'}})\big] \, \delta(q_0-E^1_p-E^{2}_{p^{\,'}})   ~~;~~ \vp^{\,\,'} = \vq-\vp\,, \label{varro1} \ee
 \be \tilde{\varrho}^{II}(q_0,\vq) = \frac{\lambda^2}{2\,\Omega_q}\,\int \frac{d^3p}{(2\pi)^3}\frac{1}{2\,E^{1}_{p}\,2\,E^{2}_{p^{\,'}}} \,   \big[ n(E^{2}_{p^{\,'}})- n(E^{1}_{p})   \big]\, \delta(E^1_p-q_0- E^{2}_{p^{\,'}})     ~~;~~ \vp^{\,\,'} = \vq-\vp  \label{varro2} \ee

\section{Spectral density for the threshold case:}\label{app:specdensthre}  For this case we consider equal masses $m_1=m_2\equiv m$, namely $E^1_{p} = E^2_p=\sqrt{p^2+m^2}$  and the threshold singularity corresponds to $M^2=4m^2$. Writing
\be E_{p^{\,'}}= \sqrt{q^2+p^2-2qp\cos(\theta)+m^2} \equiv z  \Rightarrow d(\cos(\theta)) = - \frac{zdz}{q p}\label{change}\ee  and recognizing that the delta function in (\ref{varro1}) only receives support from the region $q_0 > 0$ we find
\be \tilde{\varrho}^{I}(q_0,\vq) = \frac{\lambda^2}{32\,\pi^2 q\,\Omega_q} \, \int^\infty_{m} \big[1 + n(E_p)+n(q_0-E_p) \big]\,\int^{z^+}_{z^-} \delta(q_0-E_p-z)dz\,dE_p\,, \label{barrone} \ee where
\be z^{\pm} = \sqrt{q^2+E^2_p \pm 2q\sqrt{E^2_p-m^2}}\,,\ee
the region of support is in the domain
\be E^- \leq E \leq E^+ \,,\label{domain} \ee where
\be E^{\pm} = \frac{1}{2} \Bigg\{q_0 \pm q\,\Big[ \frac{q^2_0 - q^2 - 4m^2}{q^2_0 - q^2}\Big]^{1/2} \Bigg\}\,,     \ee therefore
\be   \tilde{\varrho}^{I}(q_0,\vq) = \frac{\lambda^2}{32\,\pi^2 q\,\Omega_q} \, \int^{E^{+}}_{E^-} \big[1 + n(E)+n(q_0-E) \big]\,dE \,. \label{rhounint}\ee

Using the identity
\be n(E) = \frac{1}{\beta} \,\frac{d}{dE} \ln[1-e^{-\beta E}]\,, \label{idn}  \ee
we find the result
\be \tilde{\varrho}^{I}(q_0,\vq) = \frac{\lambda^2}{32\,\pi^2  \,\Omega_q} \, \Bigg\{ \Big[ \frac{q^2_0 - q^2 - 4m^2}{q^2_0 - q^2}\Big]^{1/2} + \frac{2}{q\beta} \, \ln\Big[\frac{1-e^{-\beta E^+}}{1-e^{-\beta E^-}} \Big] \Bigg\}\Theta(q_0 - \sqrt{q^2+4m^2}) \,.\label{2parts}\ee
  This contribution to the spectral density arises from the two particle branch cut.

Similarly
 \be \tilde{\varrho}^{II}(q_0,\vq) =  \frac{\lambda^2}{32\,\pi^2 q\,\Omega_q} \, \int^\infty_{m} \big[n(E_p-q_0)- n(E_p)\big]  \,\int^{z^+}_{z^-} \delta(E_p-q_0-z)dz\,dE_p\,, \label{barrdos} \ee for which  we find
 \be \tilde{\varrho}^{II}(q_0,\vq) = \frac{\lambda^2}{16\,\pi^2 q\,\Omega_q \beta }\,\ln\Big[\frac{1-e^{-\beta w^+}}{1-e^{-\beta w^-}} \Big]\,\Theta(q^2-q^2_0) \label{ld}\,, \ee with

\be w^{\pm} = \frac{1}{2} \Bigg\{  q\,\Big[ \frac{q^2  - q^2_0 + 4m^2}{q^2  - q^2_0}\Big]^{1/2}\pm q_0 \Bigg\}\,.    \ee This contribution to the spectral density only has support below the light cone and vanishes at zero temperature, it does not contribute to the resonance region of the functions $C_{\mp}(q_0,t)~;~S_{\mp}(q_0,t)$.

\section{Spectral density for the infrared case:}\label{app:specdensir}
For this case we consider $\Omega_q = \sqrt{q^2+M^2}, m_1 = M, m_2 =0$, for which
 $\varrho^I(q_0,q)~;~\varrho^{II}(q_0,q)$ feature the same form as eqn. (\ref{barrone},\ref{barrdos}) but now ($E_p = \sqrt{p^2+M^2}$) and
 \be z^{+} = p+q~~;~~ z^{-} = |p-q|\,. \ee

 For $\varrho^I(q_0,q)$ the domain of support of the delta function in (\ref{barrone}) is $z^{-} \leq q_0 - E_p \leq  z^{+}$  which  is fulfilled for $q_0 > \Omega_q $ and  $E^- \leq E_p \leq E^+$ with
 \be E^\pm =  q_0 - \frac{q^2_0 - \Omega^2_q}{2(q_0 \pm q)}\,, \label{eplumin}  \ee therefore
 \be   \tilde{\varrho}^{I}(q_0,\vq) = \frac{\lambda^2}{32\,\pi^2 q\,\Omega_q} \, \int^{E^{+}}_{E^-} \big[1 + n(E)+n(q_0-E) \big]\,dE \,, \label{rhounintir}\ee
 yielding \be \tilde{\varrho}^{I}(q_0,\vq) = \frac{\lambda^2}{32\,\pi^2 \,\Omega_q} \, \Bigg\{ \Bigg[ \frac{q^2_0 - \Omega^2_q}{q^2_0 - q^2}\Bigg]  + \frac{1}{q\beta} \, \ln\Bigg[\Bigg(\frac{1-e^{-\beta E^+}}{1-e^{-\beta E^-}}\Bigg)\Bigg( \frac{1-e^{-\beta (q_0- E^-)}}{1-e^{- \beta (q_0- E^+) }} \Bigg)\Bigg]  \Bigg\}\Theta(q_0 - \Omega_q )\,.\label{2partsir}\ee
 For $\varrho^{II}(q_0,q)$ the domain of support of the delta function in (\ref{barrdos}), namely $z^{-} \leq E_p - q_0 \leq z^{+}$ with $E_p = \sqrt{p^2+M^2}$ features two different cases:

 \textbf{ Case A:} $E_p -q_0$ intersects $p+k$ or $q-p~;~ 0\leq p \leq q $ \emph{but not} $p-q~;~ p \geq q$ corresponding to the domain
 \be E_{+} \leq E_p \leq \infty \ee and
 \be -q \leq q_0 \leq q \,,\label{domaqo} \ee  therefore
 \be \tilde{\varrho}^{II}(q_0,\vq) =  \frac{\lambda^2}{32\,\pi^2 q\,\Omega_q} \, \int^\infty_{E_-} \big[n(E -q_0)- n(E )\big]  \,dE \,, \label{barrdosira} \ee
  yielding
 \be \tilde{\varrho}^{II}_A(q_0,q)  = \frac{\lambda^2}{32\,\pi^2  \,\Omega_q\,\beta\,q} \,\,\ln\Bigg[\frac{1-e^{-\beta\,E_{+}}} {1-e^{-\beta\,(E_{+}-q_0)}} \Bigg]\,\Theta(q^2-q^2_0)\,. \label{rho2caso1}  \ee This contribution is far off the resonance regions $q_0 \simeq \pm \Omega_q$.

\textbf{ Case B:} $E_p -q_0$ intersects $p+k$ or $q-p~;~ 0\leq p \leq q $ \emph{and}  $p-q~;~ p \geq q$ corresponding to the domain
 \be E_{+} \leq E_p \leq E_{-}\,, \ee and
 \be  q \leq q_0 \leq \Omega_q \,,\label{domaqo2} \ee leading to
 \be -q \leq q_0 \leq q \,,\label{domaqob} \ee  therefore
 \be \tilde{\varrho}^{II}_B(q_0,\vq) =  \frac{\lambda^2}{32\,\pi^2 q\,\Omega_q} \, \int^{E^+}_{E_-} \big[n(E -q_0)- n(E )\big]  \,dE \,, \label{barrdosirb} \ee
  yielding
\be \tilde{\varrho}^{II}_B(q_0,q)  =\frac{\lambda^2}{32\,\pi^2  \,\Omega_q\,\beta\,q} \, \left\{\ln\Bigg[\frac{1-e^{-\beta\,(E_{-}-q_0)}} {1-e^{-\beta\,(E_{+}-q_0)}} \Bigg]-\ln\Bigg[\frac{1-e^{-\beta\,E_{-}}} {1-e^{-\beta\,E_{+}}} \Bigg]\right\} \,\Theta(\Omega_q-q_0)\,\Theta(q_0-q)\,. \label{rho2casoB}  \ee This contribution   features support near the resonance region $q_0 \simeq \Omega_q$, therefore it contributes to the long time dynamics.


\begin{thebibliography}{99}

\bibitem{polko} A. Polkovnikov, K. Sengupta, A. Silva, M. Vengalattore,   Rev. Mod. Phys. 83 (2011) 863.

\bibitem{gogo} C. Gogolin and J. Eisert,  Reports on
Progress in Physics 79 no. 5, (2016) 056001.

\bibitem{weiss} U. Weiss, \emph{Quantum dissipative systems} (3rd Ed.) (World Scientific, Singapore, 2008).

\bibitem{rammer} J. Rammer, \emph{Quantum transport theory}, (Taylor and Francis, Boca Raton, 2004).

\bibitem{calzetta} E. A. Calzetta and B.-L. Hu, \emph{Nonequilibrium Quantum Field Theory}, (Cambridge University Press,
Cambridge UK, 2008).

\bibitem{bernstein} J. Bernstein, \emph{Kinetic theory in the expanding Universe}, (Cambridge monographs on Mathematical Physics, Cambridge University Press, New York, 1988).

\bibitem{kolb} E. W. Kolb, M. S. Turner, \emph{The Early Universe}, (Addison-Wesley, Reading, Massachusetts, 1990).

\bibitem{dodelson} S. Dodelson, \emph{Modern cosmology} (Academic Press, Boston, 2003).

\bibitem{raju} For a recent review see:  J. Berges, M. P. Heller, A. Mazeliauskas, R. Venugopalan,  Rev. Mod. Phys. 93, 035003 (2021), and references therein.

\bibitem{blaizot} For a review see:  J.-P. Blaizot and E. Iancu, Phys. Rept. 359, 355 (2002), and references therein.

\bibitem{biondi} S. Biondini \emph{et.al.},  Int.J.Mod.Phys. A33, 1842004  (2018).

\bibitem{irboyrai} M. Rai, L. Chen, D. Boyanovsky, Phys. Rev. D 104, 085021 (2021).


\bibitem{irthres} D. Boyanovsky, Phys. Rev. D. Phys. Rev. D 105, 056012 (2022).

\bibitem{bn} F. Bloch and A. Nordsieck, Phys. Rev.52(1937) 54.

\bibitem{lee} T.D. Lee and M. Nauenberg, Phys. Rev.133(1964) B1549.

\bibitem{chung} V. Chung, Phys. Rev.140(1965) B1110.

\bibitem{kino} T. Kinoshita,  Progress of Theoretical Physics,   5, 1045 (1950);  J. Math. Phys.3(1962) 650.

\bibitem{kibble} T.W.B. Kibble, J. Math. Phys.9(1968) 315; Phys. Rev.173(1968) 1527; Phys. Rev.174(1968) 1882; Phys. Rev.175(1968) 1624.

\bibitem{yennie}  D. R. Yennie, S. C. Frautschi, and H. Suura, , Annals Phys.13  379 (1961);  G. Grammer, Jr.,  D. R. Yennie,  Phys. Rev.D8 4332 (1973).

 \bibitem{weinberg}   S. Weinberg,  Phys. Rev.140   B516 (1965).

\bibitem{kulish}     P.P. Kulish and L.D. Faddeev, Theor. Math. Phys.4 (1971) 745.

\bibitem{greco} M. Greco, G. Rossi, Nuovo Cimento 50, 168 (1967); M. Greco, Phys. Lett. 77B, 282 (1975).

\bibitem{schwartz1}  H. Hannesdottir, M. D. Schwartz,  Phys. Rev. D 101, 105001 (2020).

 \bibitem{finites}  H. Hannesdottir, M. D. Schwartz, arXiv:1906.03271.

\bibitem{schwartz2}    C. Frye, H. Hannesdottir, N. Paul, M. D. Schwartz, K. Yan,  	Phys. Rev. D 99, 056015 (2019).

\bibitem{strominger1}  A. Strominger,  arXiv:1703.05448,

\bibitem{strominger2}  D. Kapec, M. Perry, A.-M. Raclariu, A. Strominger,  Phys. Rev. D 96, 085002 (2017).

\bibitem{kni} B. Kniehl, Nucl. Phys. B357, 439 (1991); B376, 3 (1992).

 \bibitem{will} T. Bhattacharya, S. Willenbrook, Phys. Rev. D. 47, 4022 (1993).

 \bibitem{thres1} B. A. Kniehl, C. P. Palisoc, A. Sirlin, Phys.Rev.D66, 057902 (2002); Nucl.Phys. B591, 296 (2000).

  \bibitem{thres} D. Chway, T. H. Jung, H. D. Kim,     J.Korean Phys.Soc. 69, 16 (2016).

  \bibitem{orto}  R. Schmidt, \emph{et. al}   Rep. Prog. Phys. 81 024401 (2018).

  \bibitem{orto2} M. Knap, \emph{et. al} Phys. Rev. X 2, 041020 ((2012).

   \bibitem{levitov} C. Lewandowski and L. S. Levitov, Phys. Rev. Lett.120, 076601 (2018); Phys. Rev. B97, 115423 (2018).



  \bibitem{gold} L.-Y. Chen, N. Goldenfeld and Y. Oono, Phys. Rev. Lett.73, 1311 (1994); Phys. Rev.E 54, 376 (1996).


 \bibitem{boyvega} D. Boyanovsky, H. J.de Vega,  Annals Phys. 307, 335 (2003);  D. Boyanovsky, H. J. de Vega, S. -Y. Wang,  Phys.Rev. D61, 065006 (2000);  D. Boyanovsky, H. J. de Vega, R. Holman, M. Simionato, Phys.Rev. D60, 065003  (1999).


\bibitem{dark}  D. Boyanovsky, M. Rai, L. Chen,  Phys. Rev. D 104, 123552 (2021).





\bibitem{breuer} N. P. Breuer, F. Petruccione, \emph{The theory of open quantum systems}, Oxford University Press, Oxford, 2007.

\bibitem{zoeller} C. Gardiner, P. Zoeller, \emph{Quantum Noise} Springer-Verlag, Berlin (2010).

\bibitem{lin} G. Lindblad, Comm. Math. Phys. 48, 119 (1976).

\bibitem{gori} V. Gorini, A. Kossakowski, E.C.D. Sudarshan, J. Math. Phys. 17, 821, (1976).

\bibitem{pearle} G. Pearle, European Journal of Physics 33, 805 (2012).

\bibitem{weinberg1} S. Weinberg, Phys. Rev. A90, 042102 (2014).

\bibitem{weinberg2} S. Weinberg, Phys. Rev. A93, 032124 (2016).

\bibitem{weinberg3} S. Weinberg, Phys. Rev. A94, 042117 (2016).



 \bibitem{banks} T. Banks, L. Susskind, M. Peskin, Nucl. Phys. B244, 125 (1984).

\bibitem{openburra} C. Burrage, C. Käding, P. Millington, J. Minář, Phys. Rev. D 100, 076003 (2019).

\bibitem{openaka}  Y. Akamatsu,     Prog.Part.Nucl.Phys. 123, 103932  (2022);  Phys. Rev. D 91, 056002 (2015).

\bibitem{openyao}  X. Yao,  	Int. J. of Mod. Phys. A, Vol. 36, No. 20, 2130010 (2021).

\bibitem{openmiura} Y, Akamatsu, T, Miura,     EPJ Web Conf. 258, 01006 (2022) (arXiv:2111.15402).

 \bibitem{openbram}  N. Brambilla, M. A. Escobedo, J. Soto, A. Vairo, Phys. Rev. D 96, 034021 (2017).


\bibitem{boyopen} D. Boyanovsky,  New J. Phys. 17  063017, (2015).

\bibitem{kms} R. Kubo, J. Phys. Soc. Jpn, \textbf{12}, 570 (1957);
 P. C. Martin and J. Schwinger, Phys. Rev.\textbf{ 115}, 1342
(1959).



\end{thebibliography}
\end{document}